\documentclass[english,notitlepage]{revtex4-2}
\usepackage[T1]{fontenc}
\usepackage[latin9]{inputenc}
\usepackage{babel}
\usepackage{amsmath}
\usepackage{amssymb}
\usepackage{graphicx}
\usepackage[unicode=true,
 bookmarks=false,
 breaklinks=false,pdfborder={0 0 1},backref=false,colorlinks=false]
 {hyperref}

\makeatletter

\usepackage{babel}

\usepackage{xcolor}
\usepackage{pdfcolmk}

\usepackage{ulem}

\providecolor{lyxadded}{rgb}{0,0,1}
\providecolor{lyxdeleted}{rgb}{1,0,0}

\DeclareRobustCommand{\lyxsout}[1]{\ifx\\#1\else\sout{#1}\fi}
\allowdisplaybreaks

\makeatother

\begin{document}
\title{Inclusive photoproduction of charmonia-bottomonia pairs}
\author{Marat Siddikov}
\affiliation{Departamento de Física, Universidad Técnica Federico Santa María,~~~~~~\\
 y Centro Científico - Tecnológico de Valparaíso, Casilla 110-V, Valparaíso,
Chile}
\begin{abstract}
In this preprint we analyze the inclusive photoproduction of heavy charmonia-bottomonia
pairs in the Color Glass Condensate framework and demonstrate that the cross-section of the process is sensitive to dipole and quadrupole forward scattering
amplitudes (2- and 4-point correlators of Wilson lines). Using the
phenomenological parametrizations of these amplitudes, we estimate numerically the production cross-sections in the kinematics
of the forthcoming Electron Ion Collider and the ultraperipheral collisions
at LHC. We found that the contribution controlled by the quadrupole
amplitude is dominant, and for this reason, the suggested channel
can be used as a gateway for studies of this nonperturbative object.
\end{abstract}
\maketitle

\section{Introduction}

Due to their heavy masses and small sizes, the quarkonia have been
considered as important probes of the gluonic field of the target
almost since their discovery~\cite{Korner:1991kf,Neubert:1993mb}.
The modern NRQCD framework allows to evaluate systematically various
perturbative corrections to processes which include quarkonia~\cite{Bodwin:1994jh,Maltoni:1997pt,Cho:1995ce,Cho:1995vh,Baranov:2002cf,Baranov:2007dw,Baranov:2016clx,Baranov:2015laa,Baranov:2011ib,Feng:2015cba,Brambilla:2010cs}.
This framework allows to get reasonable estimates for the cross-sections,
although at present it includes some uncertainties in nonperturbative
Long Distance Matrix Elements (LDMEs) of quarkonia states~\cite{Baranov:2016clx,Baranov:2015laa,Feng:2015cba,Brambilla:2010cs},
as well as apparent mismatch of the LDMEs of different quarkonia which
challenge the expectations based on heavy quark mass limit~\cite{Brambilla:2010cs,Han:2014jya,Shao:2014yta,Butenschoen:2014dra,Braaten:1999qk}.
Furthermore, even in the heavy quark mass limit the production of
single quarkonia provides only limited information about the gluonic
field. For this reason, since the early days of QCD the theoretical
efforts were dedicated to production mechanisms of multiple quarkonia
in the final state~\cite{Brodsky:1986ds,Lepage:1980fj,Berger:1986ii,Baek:1994kj},
which could provide a significantly more exhaustive information about
the gluonic content of the target. Recent experiments at LHC demonstrated
a feasibility to study the production of quarkonia pairs experimentally.

In this paper we will focus on the associated production of charmonia
and bottomonia, which has a very simple structure of the partonic
amplitude. The theoretical studies of such processes (particularly,
$J/\psi+\Upsilon$ hadroproduction) have been initiated in~\cite{Ko:2010xy,Likhoded:2015zna,Shao:2016wor}
and attracted a lof of theoretical attention as possible probes of
the gluonic field of the target. In the collinear and $k_{T}$ factorization
approach, such processes give a possibility to study the so-called
double parton distribution functions (DPDFs), which encode the information
about correlation of different constituents in the target. The recent
experimental data~\cite{LHCb:2023qgu,D0:2015dyx} demonstrated that
the double parton distributions might give significant contributions
in this channel. However, due to a large number of different mechanisms
in hadroproduction, these contributions have sophisticated structure,
which complicates interpretation of experimental data in terms of
partonic content of the target. At present the situation remains unclear: while inclusion of the DPDFs allows to explain the difference between
the data and predictions based on single-parton scattering,
the value of the so-called effective cross-section $\sigma_{{\rm eff}}$ (parameter which controls the magnitude of the DPDFs) depends significantly on the
channel used for its extraction~\cite{Belyaev:2017sws,Lansberg:2016rcx,Lansberg:2019adr}. In view of these difficulties of hadroproduction channel, the \textit{photo}production of the same states can serve as a simpler alternative for testing our understanding of the contributing mechanisms. Previously,
the studies of the double quarkonia photoproduction mostly focused
on exclusive channels~\cite{Goncalves:2015sfy,Goncalves:2019txs,Goncalves:2006hu,Baranov:2012vu,Yang:2020xkl,Goncalves:2016ybl,Andrade:2022rbn,Andrade:2022wph,Siddikov:2022bku,Siddikov:2023qbd}.
However, the inclusive production of such states also deserves interest
as a potential gateway for studies of the gluonic distributions. Nowadays
such processes may be studied in ultraperipheral $pA$ collisions
at the LHC, provided that the recoil nucleus is separated by large
rapidity gap from the other hadrons. The use of heavy ions allows
to boost significantly the flux of equivalent photons, and in this
way facilitates measurement of tiny cross-sections. In the future,
such processes may also be studied in electron-proton collisions at
the Electron Ion Collider (EIC)~\cite{Accardi:2012qut,AbdulKhalek:2021gbh},
the Large Hadron electron Collider (LHeC)~\cite{AbelleiraFernandez:2012cc},
and the Future Circular Collider (FCC-he)~\cite{Mangano:2017tke,Agostini:2020fmq,Abada:2019lih}.

In the kinematics of all the above-mentioned experiments, due to enhanced
role of multiparton distributions, which eventually lead to onset
of saturation effects, it is appropriate to analyze this process in
the frameworks with built-in saturation. In what follows we will use
for our studies a Color Glass Condensate (CGC) framework~\cite{GLR,McLerran:1993ni,McLerran:1993ka,McLerran:1994vd,MUQI,MV,gbw01:1,Kopeliovich:2002yv,Kopeliovich:2001ee,Gelis:2010nm,Iancu:2003uh},
which naturally incorporates the saturation effects and provides a
phenomenologically successful description of both hadron-hadron and
lepton-hadron collisions~\cite{Kovchegov:1999yj,Kovchegov:2006vj,Balitsky:2008zza,Kovchegov:2012mbw,Balitsky:2001re,Cougoulic:2019aja,Aidala:2020mzt,Ma:2014mri}.
The cross-sections of physical processes in this framework are expressed
in terms of the forward multipole scattering amplitudes ($n$-point
correlators of Wilson lines), which have probabilistic interpretation
and present important physical characteristics of the target. Technically,
these correlators generalize the multigluon distribution functions
used in collinear and $k_{T}$ factorization picture, providing correct
asymptotic behaviour in the high-energy limit. 

The paper is structured as follows. Below in Section~\ref{sec:FrameworkDerivation}
we briefly describe the main components of the CGC framework and then
present the theoretical results for the photoproduction of heavy quarkonia
pairs in the CGC approach. In Section~\ref{sec:Numer} we provide
numerical estimates using the phenomenological parametrizations of
dipole and quadrupole amplitudes. Finally, in Section~\ref{sec:Conclusions}
we draw conclusions.

\section{Theoretical framework}

\label{sec:FrameworkDerivation}Nowadays, the photoproduction processes
may be studied in electron-proton, proton-proton and proton-nuclear
collisions in ultraperipheral kinematics (UPC). The corresponding
cross-sections of these processes are related to the photoproduction
cross-section as 
\begin{equation}
\frac{d\sigma_{ep\to eM_{1}M_{2}X}}{dQ^{2}\,dy_{1}d^{2}\boldsymbol{k}_{1}^{\perp}dy_{2}d^{2}\boldsymbol{k}_{2}^{\perp}}\approx\frac{\alpha_{{\rm em}}}{\pi\,Q^{2}}\,\left(1-y+\frac{y^{2}}{2}\right)\left.\frac{d\sigma_{T}\left(\gamma+p\to M_{1}+M_{2}+X\right)}{dy_{1}d^{2}\boldsymbol{p}_{1}^{\perp}dy_{2}d^{2}\boldsymbol{p}_{2}^{\perp}}\right|_{\boldsymbol{p}_{a}^{\perp}\approx\boldsymbol{k}_{a}^{\perp}},\label{eq:LTSep-1}
\end{equation}
\begin{equation}
\frac{d\sigma_{pA\to AM_{1}M_{2}X}}{dy_{1}d^{2}\boldsymbol{k}_{1}^{\perp}dy_{2}d^{2}\boldsymbol{k}_{2}^{\perp}}=\int dn_{\gamma}\left(\omega\equiv E_{\gamma},\,\boldsymbol{q}_{\perp}\right)\,\left.\frac{d\sigma_{T}\left(\gamma+p\to M_{1}+M_{2}+X\right)}{dy_{1}d^{2}\boldsymbol{p}_{1}^{\perp}dy_{2}d^{2}\boldsymbol{p}_{2}^{\perp}}\right|_{\boldsymbol{p}_{a}^{\perp}\approx\boldsymbol{k}_{a}^{\perp}-\boldsymbol{q}^{\perp}}\label{eq:EPA_q-2-1}
\end{equation}
where in (\ref{eq:LTSep-1}) we use standard DIS notations, in which
$y$ is the inelasticity (fraction of electron energy which passes
to the photon), $q$ and $P$ are the 4-momenta of the photon and
proton, and $\left(y_{a},\boldsymbol{k}_{a}^{\perp}\right)$, with
$a=1,2$, are the rapidities and transverse momenta of the produced
quarkonia with respect to electron-proton or hadron-hadron collision
axis. The notation $d\sigma_{T}$ is used for the cross-section of
the photoproduction subprocess, induced by a transversely polarized
photon. The expression $dn_{\gamma}\left(\omega\equiv E_{\gamma},\,\boldsymbol{q}^{\perp}\right)$
in~(\ref{eq:EPA_q-2-1}) is the spectral density of the flux of equivalent
photons with energy $E_{\gamma}$ and transverse momentum $\boldsymbol{q}^{\perp}$
with respect to the nucleus, which was found explicitly in~\cite{Budnev:1975poe}.
The momenta $\boldsymbol{p}_{a}^{\perp}=\boldsymbol{k}_{a}^{\perp}-\boldsymbol{q}^{\perp}$
are the transverse parts of the quarkonia momenta with respect to
the produced photon. Due to nuclear form factor of the recoil nucleus,
the spectral density $dn_{\gamma}\left(\omega,\,\boldsymbol{q}^{\perp}\right)$
is strongly suppressed for the transverse momenta $\boldsymbol{q}^{\perp}$
larger than the inverse nuclear radius $R_{A}^{-1}$, so the average
values of $\boldsymbol{q}^{\perp}$ are quite small, $\left\langle \boldsymbol{q}_{\perp}^{2}\right\rangle \sim\left\langle Q^{2}\right\rangle \sim\left\langle R_{A}^{2}\right\rangle ^{-1}\lesssim\left(0.2\,{\rm GeV}/A^{1/3}\right)^{2}$,
and the transverse momentum dependence of the cross-sections in the
left-hand side of~(\ref{eq:EPA_q-2-1}) almost coincides with the
$p_{\perp}$-dependence of the photoproduction cross-section in the
right-hand side, and for this reason from now on we will tacitly assume
that $\boldsymbol{p}_{a}^{\perp}\approx\boldsymbol{k}_{a}^{\perp}$.

For further evaluations we need to fix the reference frame and write
out explicit light-cone momenta decompositions of the participating
hadrons. In what follows we will use notations $p_{1},\,p_{2}$ for
the 4-momenta of produced heavy quarkonia, and $P$ for the momentum
of the proton. We'll work in the reference frame in which the light-cone
expansion of these 4-vectors is given by 
\begin{align}
q & =\,\left(q^{+},\,0,\,\,\boldsymbol{0}_{\perp}\right),\quad q^{+}\approx2E_{\gamma}\label{eq:qPhoton}\\
P & =\left(\frac{m_{N}^{2}}{2P^{-}},\,P^{-},\,\,\boldsymbol{0}_{\perp}\right),\quad P^{-}=E_{p}+\sqrt{E_{p}^{2}-m_{N}^{2}}\approx2E_{p}\\
p_{a} & =\left(M_{a}^{\perp}\,e^{y_{a}}\,,\,\frac{M_{a}^{\perp}e^{-y_{a}}}{2},\,\,\boldsymbol{p}_{a}^{\perp}\right),\quad a=1,2,\label{eq:MesonLC}\\
 & M_{a}^{\perp}\equiv\sqrt{M_{a}^{2}+\left(\boldsymbol{p}_{a}^{\perp}\right)^{2}},
\end{align}
where $m_{N}$ is the mass of the nucleon, and $M_{1},\,M_{2}$ are
the masses of produced quarkonia. In what follows we will be mostly
interested in the high-energy collider kinematics $q^{+},P^{-}\gg\{Q,\,M_{a},\,m_{N}\}$,
when quarkonia are produced with relatively small transverse momenta.
In this kinematics it is possible to use eikonal picture and assume
that the plus-component $q^{+}$ of the photon light-cone momentum
is shared only between the produced quarkonia, namely
\begin{align}
 & q^{+}\approx2E_{\gamma}\approx M_{1}^{\perp}\,e^{y_{1}}+M_{2}^{\perp}\,e^{y_{2}}.\label{eq:qPlus}
\end{align}
The invariant energy $W$ of the $\gamma p$ collision and the invariant mass $M_{12}$ of the produced heavy quarkonia pair
are given by 
\begin{equation}
W^{2}\equiv s_{\gamma p}=\left(q+P\right)^{2}=-Q^{2}+m_{N}^{2}+2q\cdot P\approx-m_{N}^{2}+P^{-}\left(M_{1}^{\perp}\,e^{y_{1}}+M_{2}^{\perp}\,e^{y_{2}}\right),\label{eq:W2}
\end{equation}
and 
\begin{equation}
{\mathcal{M}}_{12}^{2}=\left(p_{1}+p_{2}\right)^{2}=M_{1}^{2}+M_{2}^{2}+2\left(M_{1}^{\perp}M_{2}^{\perp}\cosh\Delta y-\boldsymbol{p}_{1}^{\perp}\cdot\boldsymbol{p}_{2}^{\perp}\right)\label{eq:M12}
\end{equation}
respectively. The photoproduction cross-section $d\sigma_{T}$, which appears in~(\ref{eq:LTSep-1},\ref{eq:EPA_q-2-1}),
is the central quantity of interest for present study and will be
evaluated in the Color Glass Condensate approach~\cite{GLR,McLerran:1993ni,McLerran:1993ka,McLerran:1994vd,MUQI,MV,gbw01:1,Kopeliovich:2002yv,Kopeliovich:2001ee,Gelis:2010nm,Iancu:2003uh}.
In the following subsection~\ref{subsec:Derivation} we briefly remind
the main assumptions of this theoretical framework, and in subsection~\ref{subsec:Formalism}
we present final theoretical results of the evaluation. To avoid possible
soft factorization breaking final state interactions, in what follows
we will tacitly assume that the produced quarkonia are kinematically
well-separated from each other, namely that the invariant relative
velocity 
\begin{align}
 & v_{{\rm rel}}=\sqrt{1-\frac{p_{1}^{2}p_{2}^{2}}{\left(p_{1}\cdot p_{2}\right)^{2}}}=\sqrt{1-\frac{4M_{1}^{2}M_{2}^{2}}{\left({\mathcal{M}}_{12}^{2}-M_{1}^{2}-M_{2}^{2}\right)^{2}}}\label{eq:vrel}
\end{align}
is sufficiently large, 
\begin{equation}
v_{{\rm rel}}\gtrsim v_{{\rm rel}}^{({\rm cutoff})}=2\alpha_{s}\left(m_{c}\right)\approx0.7,\label{eq:constraint}
\end{equation}
or equivalently formulated in terms of the invariant mass of hadron
pairs,
\begin{equation}
{\mathcal{M}}_{12}^{2}\gtrsim\left(M_{1}+M_{2}\right)^{2}+2M_{1}M_{2}\left(\frac{1}{\sqrt{1-\left(v_{{\rm rel}}^{({\rm cutoff})}\right)^{2}}}-1\right).
\end{equation}

\subsection{High energy scattering in CGC picture}

\label{subsec:Derivation} For heavy quarkonia production, the heavy
quark mass $m_{Q}$ plays the role of the natural hard scale, which
determines the interaction strength of heavy quarks with the gluonic
field. In the heavy quark mass limit it is formally possible to develop
a systematic expansion over the strong coupling $\alpha_{s}\left(m_{Q}\right)\ll1$.
However, such expansion might be not very reliable in the small-$x$
limit, when the gluon fields are enhanced due to saturation effects
and reach values $A_{\mu}^{a}\sim1/\alpha_{s}$.  The interaction
of the partons with the ultrarelativistic target (shockwave) in this
kinematics is characterized by nearly instantaneous color exchange,
which can be described in the eikonal approximation. In this picture
the interaction of heavy quark with the target is described by a Wilson
line $U(\boldsymbol{x}_{\perp})$~\cite{McLerran:1993ni,McLerran:1993ka,McLerran:1994vd,Iancu:2003uh,MUQI,MV,Gelis:2010nm}
\begin{equation}
U\left(\boldsymbol{x}_{\perp}\right)=P\exp\left(ig\int dx^{-}A_{a}^{+}\left(x^{-},\,\boldsymbol{x}_{\perp}\right)t^{a}\right),\label{eq:Wilson}
\end{equation}
where $\boldsymbol{x}_{\perp}$ is the impact parameter of the parton,
$t_{a}$ are the ordinary color group generators of pQCD in fundamental
representation, and $A_{\mu}^{a}(x)=-\frac{1}{\nabla_{\perp}^{2}}\rho_{a}(x^{-},\,\boldsymbol{x})$
is the gluonic field in the target created by the color charge density
$\rho_{a}$. According to classical CGC picture~\cite{McLerran:1993ni,McLerran:1993ka,McLerran:1994vd},
for multiparton scattering processes each physical observable $\mathcal{O}$
should be averaged with a weight function $W[\rho]$ which describes
the probability of a given charge distribution $\rho$ inside the
target, namely 
\begin{equation}
\left\langle \mathcal{O}\right\rangle =\int\mathcal{D}\rho\,W[\rho]\,\mathcal{O}[\rho]\label{eq:Ave}
\end{equation}
where from now on we will use angular brackets $\langle...\rangle$
for such averaging. The analytic evaluation of the path integral $\mathcal{D}\rho$
is possible only for a few simple forms of $W[\rho]$ (e.g. for gaussian).
Fortunately, for many observables the averages $\left\langle \mathcal{O}\right\rangle $
may be expressed in terms of a limited set of universal (process-independent)
amplitudes, such as dipole or quadrupole forward scattering amplitudes,
and thus avoiding explicit evaluation of the integral over all possible
charge configurations. 

For example, the amplitude of the heavy quark pair or (single) quarkonia
production from gluon may be described by the two diagrams shown
in the Figure~(see for details~\cite{Ayala:2017rmh,Caucal:2021ent,Caucal:2022ulg}
) and is given by 
\begin{equation}
\mathcal{A}^{a}=-ig\int d^{4}z\,\bar{u}\left(p_{q},\,z\right)\gamma^{\mu}t^{b}\varepsilon_{\mu}^{ab}\left(k,z\right)v\left(p_{\bar{q}},\,z\right),\label{eq:amp_pert}
\end{equation}
where $z$ represents the coordinate of the interaction point in the
configuration space, and the corresponding parton fields are given
by

\begin{figure}
\includegraphics[width=5cm]{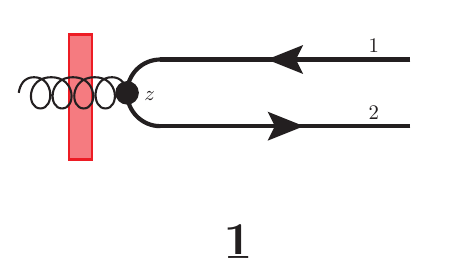}\includegraphics[width=5cm]{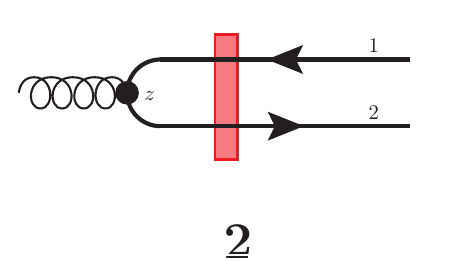}

\caption{\label{fig:CGCBasic}Diagrams which describe the inclusive heavy quark
pair production in CGC picture in the leading order over $\alpha_{s}$.
The subscript numbers 1,2 enumerate the heavy quark lines in our convention.
The subscript letters $z$ is the coordinate of the interaction vertex
in the configuration space. The red block represents the interaction
with the target (shockwave).}
\end{figure}

\begin{align}
\bar{u}\left(p_{q},\,z\right) & =\frac{1}{2}\left(\frac{p_{q}^{+}}{2\pi}\right)\int d^{2}\boldsymbol{x}_{q}e^{ip_{q}^{+}\left(z^{-}-\frac{\left(\boldsymbol{x}_{q}-\boldsymbol{z}\right)^{2}}{2z^{+}}\right)-i\boldsymbol{p}_{q}\cdot\boldsymbol{x}_{q}+\frac{iz^{+}}{2p_{q}^{+}}m^{2}}\times\label{eq:SWubar}\\
 & \times\left(\frac{i}{z^{+}}\right)\bar{u}_{p}\gamma^{+}\left[U\left(\boldsymbol{x}_{q}\right)\theta\left(-z^{+}\right)+\theta\left(z^{+}\right)\right]\left(\gamma^{-}-\frac{\hat{\boldsymbol{x}}_{q}-\hat{\boldsymbol{z}}}{z^{+}}+\frac{m}{p_{q}^{+}}\right),\nonumber 
\end{align}
for the produced quarks, 
\begin{align}
v\left(p_{\bar{q}},\,z\right) & =\frac{1}{2}\left(\frac{p_{\bar{q}}^{+}}{2\pi}\right)\int d^{2}\boldsymbol{x}_{\bar{q}}e^{ip_{\bar{q}}^{+}\left(z^{-}-\frac{\left(\boldsymbol{x}_{\bar{q}}-\boldsymbol{z}\right)^{2}}{2z^{+}}\right)-i\boldsymbol{p}_{\bar{q}}\cdot\boldsymbol{x}_{\bar{q}}+\frac{iz^{+}}{2p_{\bar{q}}^{+}}m^{2}}\times\label{eq:SWv}\\
 & \times\left(\frac{i}{z^{+}}\right)\left(\gamma^{-}-\frac{\hat{\boldsymbol{x}}_{\bar{q}}-\hat{\boldsymbol{z}}}{z^{+}}-\frac{m}{p_{\bar{q}}^{+}}\right)\left[U^{\dagger}\left(\boldsymbol{x}_{\bar{q}}\right)\theta\left(-z^{+}\right)+\theta\left(z^{+}\right)\right]\gamma^{+}v_{p_{\bar{q}}}\nonumber 
\end{align}
for produced antiquark, and
\begin{align}
\varepsilon_{\mu}^{ba}\left(k,z\right) & =\frac{k^{+}}{2\pi}\int d^{2}\boldsymbol{x}_{g}e^{-ik^{+}\left(z^{-}-\frac{\left(\boldsymbol{x}_{g}-\boldsymbol{z}\right)^{2}}{2z^{+}}\right)+i\boldsymbol{k}\cdot\boldsymbol{x}_{g}}\times\label{SWg}\\
 & \times\left(\frac{-i}{z^{+}}\right)\left(g_{\mu\sigma}^{\perp}+\frac{x_{g}^{\sigma}-z^{\sigma}}{z^{+}}n_{2}^{\mu}\right)\left[\mathcal{U}^{ba}\left(\boldsymbol{x}_{g}\right)\theta\left(z^{+}\right)+\delta^{ab}\theta\left(-z^{+}\right)\right]\varepsilon_{\perp}^{\sigma}(\boldsymbol{k})\nonumber 
\end{align}
for the initial-state gluon, where the vector $n_{2}$ in~(\ref{SWg})
is the unit light-cone vector pointing in the direction of the target
momentum. The notation $\mathcal{U}^{ba}$ is used for the components
of the matrix~(\ref{eq:Wilson}) in adjoint representation. The expressions
in square brackets in (\ref{eq:SWubar}-\ref{SWg}) merely reflect
that corresponding $U$-matrices should be taken into account only
if the colored parton exists at the moment of interaction with the
shockwave of the target. In the limit $U\to1$ (no interaction) the
amplitude~(\ref{eq:amp_pert}) reduces to a mere perturbative spinors
and gluon polarization vector. In what follows we will work in the
mixed representation, describing the kinematics of each parton in
terms of its light-cone momenta of the partons together with transverse
coordinates in configuration space. Since in the eikonal approximation
the interaction of any parton with the target is described by multiplicative
factor $U(\boldsymbol{X}_{p})$ in proper representation of the color
group, evaluation of the amplitudes and cross-sections becomes straightforward.
For a simple process shown in~(\ref{fig:CGCBasic}), the amplitude
of the process is described by the $S$-matrix element~\cite{Gelis:2010nm,Kovchegov:2012mbw,Iancu:2003uh}

\begin{equation}
S_{2}\left(Y,\,\boldsymbol{x}_{q},\,\boldsymbol{x}_{\bar{q}}\right)=\frac{1}{N_{c}}\left\langle {\rm tr}\left(U\left(\boldsymbol{\boldsymbol{x}}_{q}\right)U^{\dagger}\left(\boldsymbol{x}_{\bar{q}}\right)\right)\right\rangle _{Y},\label{eq:S_matrix}
\end{equation}
where the notation $Y$ is used for the dipole rapidity. The dipole
scattering amplitude $N(x,\,\boldsymbol{r},\,\boldsymbol{b})$ can
be related to $S_{2}\left(Y,\,\boldsymbol{x}_{q},\,\boldsymbol{x}_{\bar{q}}\right)$
as 
\begin{equation}
N\left(x,\,\boldsymbol{r},\,\boldsymbol{b}\right)=1-S_{2}\left(Y,\,\boldsymbol{x}_{q},\,\boldsymbol{x}_{\bar{q}}\right),\label{eq:NS}
\end{equation}
where the variable $\boldsymbol{r}\equiv\boldsymbol{x}_{q}-\boldsymbol{x}_{\bar{q}}$
is the transverse size of the dipole, and $\boldsymbol{b}\equiv\alpha_{q}\,\boldsymbol{x}_{q}+\alpha_{\bar{q}}\boldsymbol{x}_{\bar{q}}$
is the transverse position of the dipole's center of mass. In complete
analogy we may introduce the multipole scattering $S$-matrix elements
(correlator of Wilson lines)
\begin{equation}
S_{2n}(Y,\,\boldsymbol{x}_{1},\,\boldsymbol{\xi}_{1},...,\boldsymbol{x}_{n},\,\boldsymbol{\xi}_{n})=\frac{1}{N_{c}}\left\langle {\rm tr}\left(U\left(\boldsymbol{\boldsymbol{x}}_{1}\right)U^{\dagger}\left(\boldsymbol{\xi}_{1}\right)...U\left(\boldsymbol{\boldsymbol{x}}_{n}\right)U^{\dagger}\left(\boldsymbol{\xi}_{n}\right)\right)\right\rangle _{Y},\label{eq:MultipoleDfinition}
\end{equation}
which characterize scattering amplitudes of $2n$-particle ensemble
of quarks. Each such correlator represents independent characteristics
of the target and characterizes gluon distributions in the target.

\subsubsection{Dilute scattering limit}

\label{subsec:Dilute}If the saturation effects are not very large,
the interaction of heavy quarks with the target can be described
perturbatively, making a Taylor expansion of the Wilson line~(\ref{eq:Wilson}),
\begin{align}
U\left(\boldsymbol{x}_{\perp}\right) & \approx1+ig\int dx^{-}A_{a}^{+}\left(x^{-},\,\boldsymbol{x}_{\perp}\right)t^{a}+\label{eq:Wilson-1}\\
 & +\frac{\left(ig\right)^{2}}{2!}P\,\int dx_{1}^{-}A_{a_{1}}^{+}\left(x_{1}^{-},\,\boldsymbol{x}_{1\perp}\right)\int dx_{2}^{-}A_{a_{2}}^{+}\left(x_{2}^{-},\,\boldsymbol{x}_{2\perp}\right)t^{a_{1}}t^{a_{2}}+...\nonumber 
\end{align}
The dipole scattering amplitude~(\ref{eq:NS}) in this limit may
be expressed as

\begin{equation}
N\left(x,\,\boldsymbol{r},\,\boldsymbol{b}\right)\approx\frac{1}{4N_{c}}\left\langle \left[\gamma_{a}\left(\boldsymbol{x}_{Q}\right)-\gamma_{a}\left(\boldsymbol{x}_{\bar{Q}}\right)\right]^{2}\right\rangle +\mathcal{O}\left(\alpha_{s}\right),\label{eq:N_dip-1}
\end{equation}
where we defined $\gamma_{a}$as
\begin{equation}
\gamma_{a}(\boldsymbol{x})=g\int dx^{-}A_{a}^{+}(x^{-},\,\boldsymbol{x})=g^{2}\int dx^{-}\frac{1}{\nabla_{\perp}^{2}}\rho_{a}(x^{-},\,\boldsymbol{x}).\label{eq:gamma}
\end{equation}
For further evaluations it is convenient to rewrite~(\ref{eq:gamma})
in the form 
\begin{equation}
\frac{1}{N_{c}}\left\langle \gamma_{a}\left(\boldsymbol{x}_{1}\right)\gamma_{a}\left(\boldsymbol{x}_{2}\right)\right\rangle =-2N\left(x,\,\boldsymbol{r}_{12},\,\boldsymbol{b}_{12}\right)+\frac{\Gamma\left(\boldsymbol{x}_{1}\right)+\Gamma\left(\boldsymbol{x}_{2}\right)}{2N_{c}},\label{eq:N_dip2}
\end{equation}
where we introduced a shorthand notation $\Gamma\left(\boldsymbol{x}\right)\equiv\left|\gamma_{a}(\boldsymbol{x})\right|^{2}$,
and $\boldsymbol{r}_{12}$, $\boldsymbol{b}_{12}$ are the transverse
size and impact parameter of the color dipole. For certain processes
which involve partonic ensembles with net zero color charge, the contributions
$\sim\Gamma\left(\boldsymbol{x}_{i}\right)$ cancel, so the cross-sections
eventually can be represented entirely as a linear superposition of
the dipole amplitudes $N\left(x,\,\boldsymbol{r},\,\boldsymbol{b}\right)$. The multipole scattering amplitudes~(\ref{eq:MultipoleDfinition})
in this limit may be rewritten as
\begin{equation}
S_{2n}\left(Y,\,\boldsymbol{x}_{1},\,\boldsymbol{x}_{2},...,\boldsymbol{x}_{2n-1},\,\boldsymbol{x}_{2n}\right)\approx\sum_{aijk\ell}c_{ijk\ell}\left(\gamma_{a}\left(\boldsymbol{x}_{i}\right)-\gamma_{a}\left(\boldsymbol{x}_{j}\right)\right)\left(\gamma_{a}\left(\boldsymbol{x}_{k}\right)-\gamma_{a}\left(\boldsymbol{x}_{\ell}\right)\right),\label{eq:SMatr4}
\end{equation}
where $c_{ijk\ell}$ are some numerical coefficients. Using the identities~(\ref{eq:N_dip2}),
it is possible to express~(\ref{eq:SMatr4}) as a linear superposition
of the forward color dipole amplitudes. However, at higher orders
in $\alpha_{s}(m_{Q})$ there are contributions which include multiple
products of $\gamma\left(x_{k}\right)$, and which cannot be reduced
to a mere superposition of dipole amplitudes, demonstrating independence
of the multipole contributions in general case.

\subsection{Inclusive photoproduction of meson pairs}

\label{subsec:Formalism} 
\begin{figure}
\includegraphics[width=8cm]{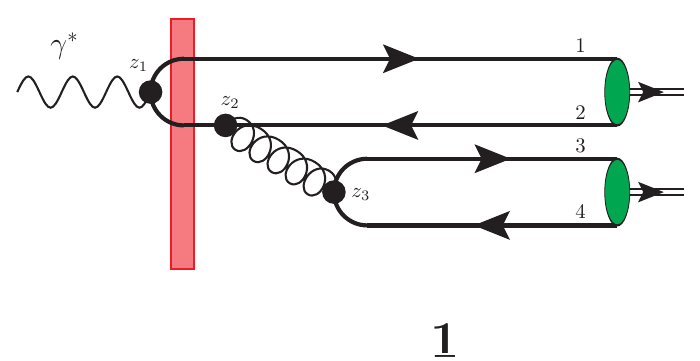}\includegraphics[width=8cm]{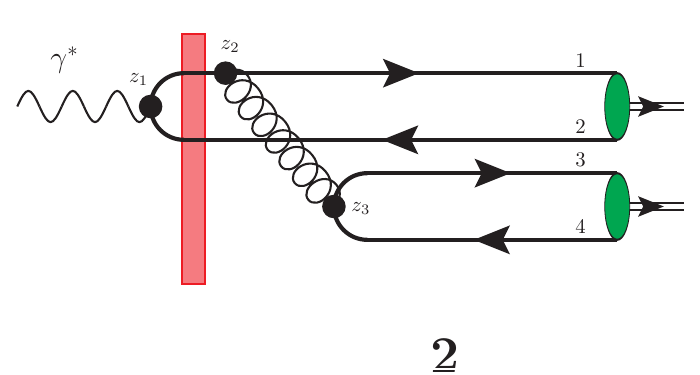}

\includegraphics[width=8cm]{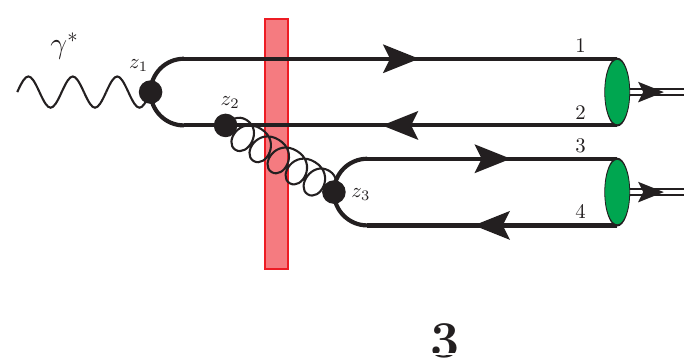}\includegraphics[width=8cm]{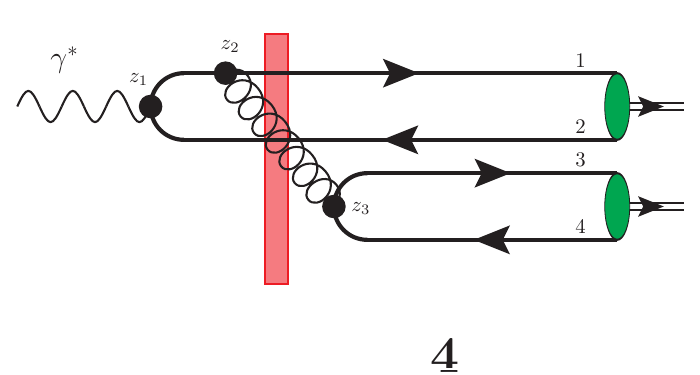}

\includegraphics[width=8cm]{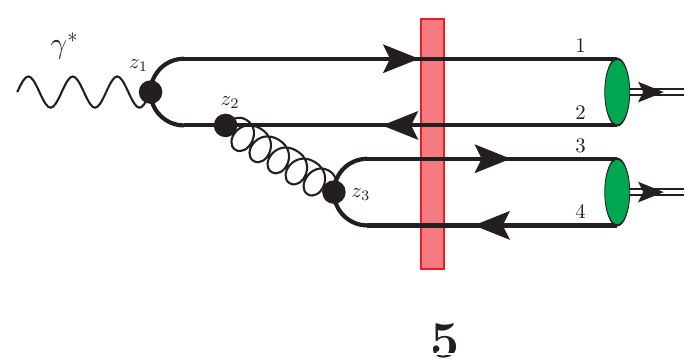}\includegraphics[width=8cm]{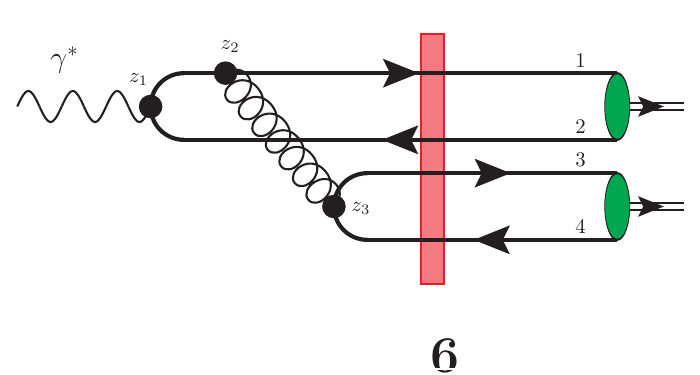}

\caption{\label{fig:QuarkoniaPairCGC}Diagrams which describe the inclusive
charmonia-bottomonia pairs production in CGC picture in the leading
order over $\alpha_{s}$. The diagrams in the right column differ
from diagrams in the left column just by inversion of the quark line
in the upper loop (charge conjugation). The dominant color singlet
contribution comes from the diagrams in the last row (diagrams \textquotedblleft 5\textquotedblright{}
and \textquotedblleft 6\textquotedblright{} respectively). The subscript
numbers 1-4 enumerate the heavy quark lines in our convention. The
subscript letters $z_{1}\,...\,z_{3}$ stand for the coordinates of
the interaction vertices in the configuration space. The red block
represents the interaction with the target (shockwave).}
\end{figure}

\begin{figure}
\includegraphics[width=5cm]{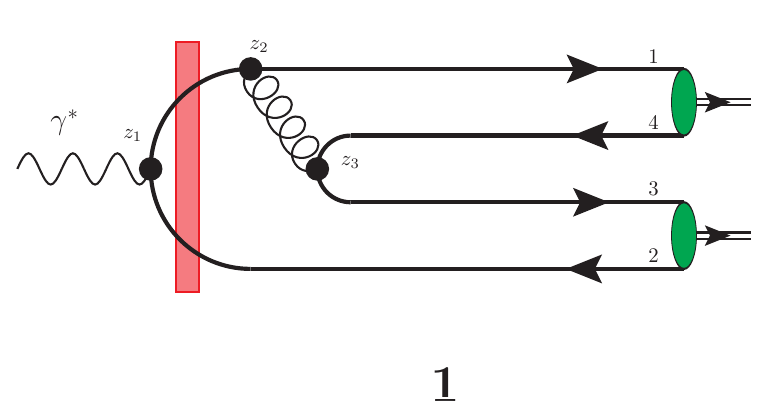}\includegraphics[width=5cm]{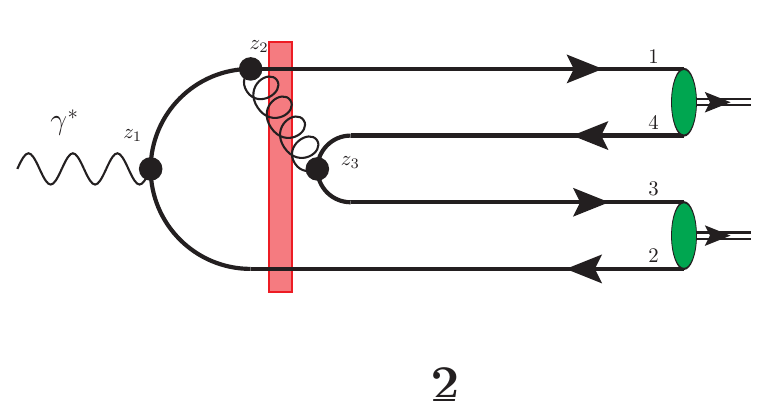}\includegraphics[width=5cm]{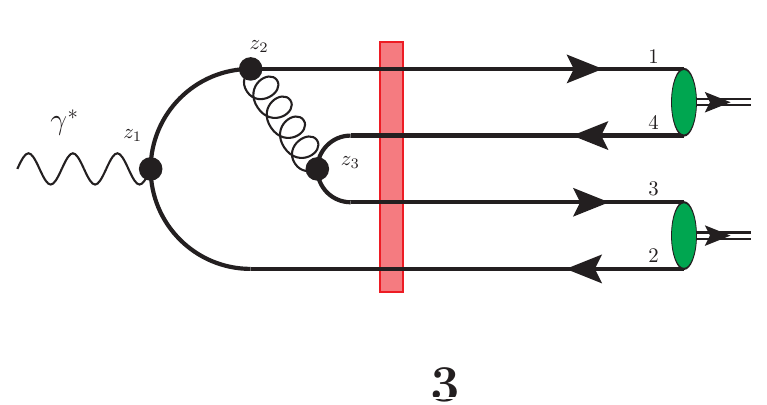}

\includegraphics[width=5cm]{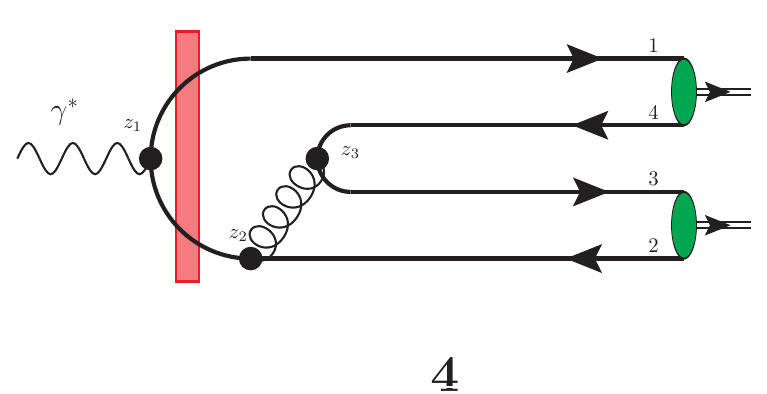}\includegraphics[width=5cm]{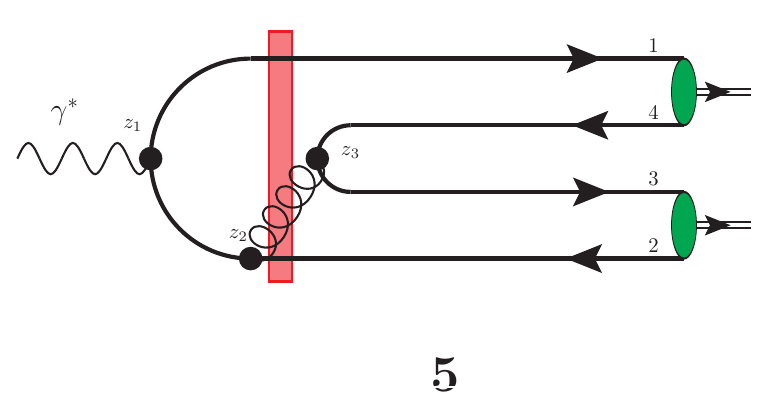}\includegraphics[width=5cm]{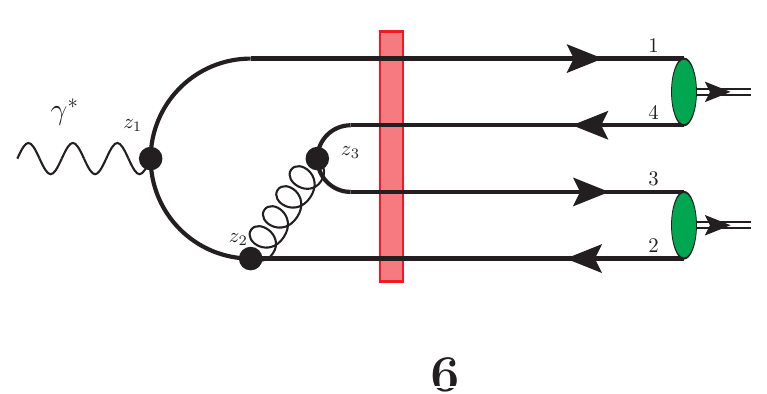}

\caption{\label{fig:QuarkoniaPairCGC-1}Additional diagrams which in general
should be taken into account for inclusive quarkonia pair production
in CGC picture in the leading order over $\alpha_{s}$, yet do not
contribute to charmonia-bottomonia pairs. The subscript numbers 1-4
enumerate the heavy quark lines in our convention. The subscript letters
$z_{1}...z_{3}$ stand for the coordinates of the interaction vertices
in the configuration space. The red block represents the interaction
with the target (shockwave).}
\end{figure}

The inclusive heavy quarkonia pair production
in general may proceed via a number of different mechanisms, and for
this reason presents a sophisticated problem. In what follows we will
focus on the production of charmonia-bottomonia pairs with opposite
$C$-parity, in order to have vacuum quantum numbers in the $t$-channel.
In the leading order of the perturbation theory at the amplitude level
this process is described by a set of diagrams shown in the Figure~\ref{fig:QuarkoniaPairCGC}.
For comparison, the production of quarkonia pairs in general case
requires inclusion of additional Feynman diagrams shown in the Figure~\ref{fig:QuarkoniaPairCGC-1},
or emission of additional hard gluons, leading to significantly more
complicated expressions for the cross-sections.

The evaluation of the diagrams shown in the Figure~\ref{fig:QuarkoniaPairCGC}
is straightforward. The diagrams in the left and right rows might
be related to each other by inversion of the quark line in the upper
loop, and for this reason for quarkonia states with definite $C$-parity
give the same (up to a sign) contributions. According to NRQCD, the
quarkonia formation can proceed both from color singlet and color
octet $Q\bar{Q}$ pairs, so the total cross-section may be represented
as an incoherent sum
\begin{equation}
\frac{d\sigma}{d\Omega_{h}}=\frac{d\sigma^{(1)}}{d\Omega_{h}}+\frac{d\sigma^{(8)}}{d\Omega_{h}},
\end{equation}
where we introduced a shorthand notation $d\Omega_{h}=dy_{1}\,d\left|p_{1}^{\perp}\right|^{2}dy_{2}\,d\left|p_{2}^{\perp}\right|^{2}d\phi$
for the phase volume, $\phi$ is the azimuthal angle between the transverse
momenta of the quarkonia, and $d\sigma^{(1)},\,\,d\sigma^{(8)}$ stand
respectively for the color singlet and octet contributions. The diagrams
1-4 lead to formation of $\bar{Q}Q$ pair (partons ``3'', ''4'')
in color octet state,  whereas the remaining two diagrams lead to
formation of both color singlet and color octet pairs. The evaluation
of the color octet contribution, shown in diagrams 1-4 in the upper
two rows may be interpreted as a production of the first quarkonium
$M_{1}$ via $\gamma^{*}\to M_{1}g$ subprocess, with subsequent fragmentation
of the emitted hard gluon into the second quarkonium $M_{2}$. Indeed,
we may observe that the gluon connecting quark lines of different
heavy flavors is perturbative, and there is no instantaneous contributions
to its propagator. For this reason, the contribution of diagrams
1-4 to the cross-section may be rewritten as 
\begin{equation}
\frac{d\sigma^{(8)}}{d\Omega_{h}}=\int\frac{dz}{z}\,D_{g\to M_{2}}\left(z\right)\frac{d\sigma_{\gamma^{*}\to M_{1}g}\left(\boldsymbol{p}_{1M},\,\boldsymbol{p}_{g}=\boldsymbol{p}_{2M}/z\right)}{d\Omega_{h}}
\end{equation}
where $D_{g\to M_{2}}\left(z\right)$ is the fragmentation function
of the gluon to the second quarkonium $M_{2}$~\cite{Baranov:2021gil,Braaten:1993rw},
and $z$ is the fraction of the gluon momentum carried by the second
quarkonium. In the leading order over $\alpha_{s}\left(m_{Q}\right)$,
the fragmentation function is given by 
\begin{equation}
D_{g\to M_{2}}\left(z\right)\approx\frac{\alpha_{s}}{m_{c}^{3}}\frac{\pi}{24}\left\langle \mathcal{O}^{M_{2}}\left(^{3}S_{1}^{[8]}\right)\right\rangle \delta\left(z-1\right)\left(1+\mathcal{O}\left(\frac{\Lambda_{{\rm QCD}}}{m_{Q}}\right)\right),
\end{equation}
where $\left\langle \mathcal{O}^{M_{2}}\left(^{3}S_{1}^{[8]}\right)\right\rangle $
is the corresponding color octet matrix element for the quarkonium
$M_{2}$, however at higher orders in $\alpha_{s}\left(m_{Q}\right)$,
due to emission of soft gluons, the function $D_{g\to M_{2}}$ acquires
nonzero finite width, and its peak shifts towards smaller $z<1$~\cite{Braaten:1993rw}.
The evaluation of the cross-section $d\sigma_{\gamma^{*}\to M_{1}g}$
may be found in~\cite{Kang:2023doo} and largely repeats similar
evaluation of the $\gamma^{*}\to\bar{Q}Qg$ subprocesses from~\cite{Ayala:2016lhd,Ayala:2017rmh}.
The final result of this evaluation reads as~\cite{Ayala:2016lhd,Ayala:2017rmh}

\begin{equation}
\frac{d\sigma_{\gamma^{*}\to M_{1}g}\left(\boldsymbol{p}_{1M},\boldsymbol{p}_{g}\right)}{d\Omega_{h}^{*}}=\sum_{i,j=1}^{4}\frac{\mathrm{d}\hat{\sigma}_{i,j}^{\lambda}}{d\Omega_{h}^{*}}
\end{equation}

\begin{align}
\frac{\mathrm{d}\hat{\sigma}_{i,j}^{\lambda}}{d\Omega_{h}^{*}} & =\frac{1}{\left(q^{+}\right)^{2}}\frac{1}{(2\pi)^{4}}\frac{1}{8(1-x_{g})}\int\mathrm{d}\Pi_{i}(\boldsymbol{p},\boldsymbol{p}_{g};\boldsymbol{r},\boldsymbol{b},\boldsymbol{z})\mathrm{d}\Pi_{j}^{\dagger}(\boldsymbol{p},\boldsymbol{p}_{g};\boldsymbol{r}^{\,\prime},\boldsymbol{b}_{12}^{\,\prime},\boldsymbol{z}^{\,\prime})\times\\
 & \times\Xi_{i,j}^{[c]}\left(\boldsymbol{b}_{12}+\frac{\boldsymbol{r}_{12}}{2},\boldsymbol{b}_{12}-\frac{\boldsymbol{r}_{12}}{2},\,\boldsymbol{z};\,\boldsymbol{b}_{12}^{\,\prime}-\frac{\boldsymbol{r}_{12}^{\,\prime}}{2},\boldsymbol{b}_{12}^{\,\prime}+\frac{\boldsymbol{r}_{12}^{\,\prime}}{2},\boldsymbol{z}^{\,\prime},Y\right)\Gamma_{i,j}^{\lambda}\left(\boldsymbol{p},\boldsymbol{p}_{g},Q;\boldsymbol{r}_{12},\boldsymbol{b}_{12},\boldsymbol{z},\boldsymbol{r}_{12}^{\,\prime},\boldsymbol{b}_{12}^{\,\prime},\boldsymbol{z}^{\,\prime}\right),\nonumber 
\end{align}
where $\boldsymbol{b}_{12}=\left(\boldsymbol{x}_{1}+\boldsymbol{x}_{2}\right)/2$
is the impact parameter of the quarkonium $M_{1}$ , $\boldsymbol{r}=\boldsymbol{x}_{1}-\boldsymbol{x}_{2}$
is the transverse distance between the quarks in the quarkonium, ``primed''
variables $\boldsymbol{b}',\boldsymbol{z}',\boldsymbol{r}'$ correspond
to similarly defined variables in the conjugate amplitude, and the
corresponding differential measure of integration $\mathrm{d}\Pi_{i}(...)$
is defined as 
\begin{align}
\mathrm{d}\Pi_{i}(\boldsymbol{p},\boldsymbol{p}_{g};\boldsymbol{r},\boldsymbol{b},\boldsymbol{z})=\mathrm{d}^{2}\boldsymbol{r}\;\;e^{-i{\bf p}_{1}\cdot{\bf b}}\mathrm{d}^{2}\boldsymbol{b}\begin{cases}
 & \mathrm{d}^{2}\boldsymbol{z}e^{-i\boldsymbol{p}_{g}\cdot{\bf z}}\quad\text{ if }\quad i=1,3\\
 & e^{-i\boldsymbol{p}_{g}\cdot\left({\bf b}+\frac{{\bf r}}{2}\right)}\quad\text{ if }\quad i=2\;\\
 & e^{-i\boldsymbol{p}_{g}\cdot\left({\bf b}-\frac{{\bf r}}{2}\right)}\quad\text{ if }\quad i=4\;,
\end{cases}
\end{align}

The explicit expressions for the matrices $\Xi_{i,\,j}^{[c]}$ and
$\Gamma_{i,\,j}^{\lambda}$ may be found in~\cite{Kang:2023doo}
and will be omitted here for brevity. According to modern estimates~\cite{Feng:2015cba,Baranov:2019lhm,Baranov:2021gil,Feng:2015wka,Abdulov:2020nxh,Baranov:2016clx},
the color octet matrix elements $\mathcal{O}^{M}\left(^{3}S_{1}^{[8]}\right)$
are very small and constitute $\lesssim10^{-2}$ of the color singlet
matrix elements. For this reason, a gluon fragmentation mechanism
constitutes just a few percent correction and may be disregarded
in the first approximation, especially in the kinematics of small
transverse moments $p^{\perp}$ which we study in this paper. 

In what follows we will focus on the contribution of the last two
diagrams 5, 6 in the Figure~\ref{fig:QuarkoniaPairCGC} which represent
the dominant color singlet contributions. In evaluations we may use
the fact that the final-state quarks are nearly onshell, and in eikonal
picture the free quark propagator becomes diagonal in transverse coordinates
in configuration space, namely 
\begin{equation}
S_{0}\left(x-y\right)\approx\int\frac{dk^{+}d^{2}\boldsymbol{k}_{\perp}}{\left(2\pi\right)^{3}}\frac{\gamma^{+}}{2k^{+}}e^{-ik\cdot(x-y)}\sim\delta\left(\boldsymbol{x}_{\perp}-\boldsymbol{y}_{\perp}\right),
\end{equation}
(see details in~\cite{Belitsky:2002sm}). The corresponding cross-section
in this limit may be rewritten as a convolution of wave functions
and target-dependent multipole contributions, namely: 
\begin{align}
\frac{d\sigma}{d\Omega} & =\int\left(\prod_{i=1}^{4}d^{2}\boldsymbol{x}_{i}\right)\left(\prod_{i=1}^{4}d^{2}\boldsymbol{x}_{i}'\right)\mathcal{N}\left(\boldsymbol{x}_{1},\,\boldsymbol{x}_{2},\,\boldsymbol{x}_{3},\,\boldsymbol{x}_{4};\,\boldsymbol{x}_{1}',\,\boldsymbol{x}_{2}',\,\boldsymbol{x}_{3}',\,\boldsymbol{x}_{4}'\right)\times\label{eq:XSec}\\
 & \times\Psi_{M_{1}}^{\dagger h_{1}h_{2}}\Psi_{M_{2}}^{\dagger h_{3}h_{4}}\psi_{\bar{Q}Q\bar{Q}Q}^{(\gamma)h_{1}h_{2}h_{3}h_{4}}\left(\boldsymbol{x}_{1};\,\boldsymbol{x}_{2};\,\boldsymbol{x}_{3};\,\boldsymbol{x}_{4};\,q\right)e^{i\left(\boldsymbol{p}_{1}^{\perp}\cdot\boldsymbol{b}_{12}+\boldsymbol{p}_{2}^{\perp}\cdot\boldsymbol{b}_{34}\right)}\times\nonumber \\
 & \times\left(\Psi_{M_{1}}^{\dagger\mathfrak{h}_{1}\mathfrak{h}_{2}}\Psi_{M_{2}}^{\dagger\mathfrak{h}_{3}\mathfrak{h}_{4}}\psi_{\bar{Q}Q\bar{Q}Q}^{(\gamma)\mathfrak{h}_{1}\mathfrak{h}_{2}\mathfrak{h}_{3}\mathfrak{h}_{4}}\left(\boldsymbol{x}_{1}';\,\boldsymbol{x}_{2}';\,\boldsymbol{x}_{3}';\,\boldsymbol{x}_{4}';\,q\right)e^{i\left(\boldsymbol{p}_{1}^{\perp}\cdot\boldsymbol{b}_{12}'+\boldsymbol{p}_{2}^{\perp}\cdot\boldsymbol{b}_{34}'\right)}\right)^{*},\nonumber
\end{align}
where $h_{i}$ and $\mathfrak{h_{i}}$ are the helicities of heavy
quarks in the amplitude and its conjugate (summation is implied),
$\psi_{\bar{Q}Q\bar{Q}Q}^{(\gamma)h_{1}h_{2}h_{3}h_{4}}\left(\boldsymbol{x}_{1};\,\boldsymbol{x}_{2};\,\boldsymbol{x}_{3};\,\boldsymbol{x}_{4};\,q\right)$
is the wave function which describes the 4-quark $\bar{Q}Q\bar{Q}Q$
Fock state inside the quasireal photon~(see details in Appendix~\ref{sec:WF}).
We also use a shorthand notation $\Psi_{M}^{h_{i}h_{j}}$ for the
conventional NRQCD projectors (in helicity basis) multiplied by the
appropriate Long Distance Matrix Elements (LDMEs) of the corresponding
meson $M$. In the heavy quark mass limit the largest LDMEs are associated
with the $^{3}S_{1},\,{}^{1}S_{0}$ and $^{1}P_{0}$ states, for which
the corresponding projectors are given by~\cite{Cho:1995ce,Cho:1995vh,DVMPcc1,Baranov:2007dw}
\begin{align}
\Psi^{\dagger h_{1}h_{2}}\left[^{1}S_{0}\right] & =-\frac{1}{4N_{c}}\sqrt{\frac{\left\langle \mathcal{O}_{M}\left[^{1}S_{0}\right]\right\rangle }{m_{Q}}}\,\left[\gamma_{5}\left(\frac{\hat{P}_{M}}{2}+m_{Q}\right)\right]_{h_{1}h_{2}},\label{eq:S1}\\
\Psi^{\dagger h_{1}h_{2}}\left[^{3}S_{1}\right] & =-\frac{1}{4N_{c}}\sqrt{\frac{\left\langle \mathcal{O}_{M}\left[^{1}S_{0}\right]\right\rangle }{m_{Q}}}\,\left[\hat{\varepsilon}\left(S_{z}\right)\left(\frac{\hat{P}_{M}}{2}+m_{Q}\right)\right]_{h_{1}h_{2}},\label{eq:S3}
\end{align}
where $P_{M}$ is the 4-momentum of the corresponding quarkonium,
$m_{Q}$ are the quarkonia masses, $\left\langle \mathcal{O}_{M}\right\rangle $
are the corresponding long-distance matrix elements. We also introduced
a shorthand notation
\begin{align}
 & \mathcal{N}\left(\boldsymbol{x}_{1},\,\boldsymbol{x}_{2},\,\boldsymbol{x}_{3},\,\boldsymbol{x}_{4};\,\boldsymbol{x}_{1}',\,\boldsymbol{x}_{2}',\,\boldsymbol{x}_{3}',\,\boldsymbol{x}_{4}'\right)=\label{eq:NDef}\\
 & \left\langle {\rm tr}_{c}\left[t_{a}U\left(\boldsymbol{x}_{1}\right)U^{\dagger}\left(\boldsymbol{x}_{2}\right)\right]{\rm tr}_{c}\left[t_{a}U\left(\boldsymbol{x}_{3}\right)U^{\dagger}\left(\boldsymbol{x}_{4}\right)\right]{\rm tr}_{c}\left[t_{b}U\left(\boldsymbol{x}_{1}'\right)U^{\dagger}\left(\boldsymbol{x}_{2}'\right)\right]{\rm tr}_{c}\left[t_{b}U\left(\boldsymbol{x}_{3}'\right)U^{\dagger}\left(\boldsymbol{x}_{4}'\right)\right]\right\rangle =\nonumber \\
 & \frac{1}{4}\left\langle \left({\rm tr}_{c}\left[U\left(\boldsymbol{x}_{1}\right)U^{\dagger}\left(\boldsymbol{x}_{2}\right)U\left(\boldsymbol{x}_{3}\right)U^{\dagger}\left(\boldsymbol{x}_{4}\right)\right]-\frac{1}{N_{c}}{\rm tr}_{c}\left[U\left(\boldsymbol{x}_{1}\right)U^{\dagger}\left(\boldsymbol{x}_{2}\right)\right]{\rm tr}_{c}\left[U\left(\boldsymbol{x}_{3}\right)U^{\dagger}\left(\boldsymbol{x}_{4}\right)\right]\right)\right.\times\nonumber \\
 & \times\left.\left({\rm tr}_{c}\left[U\left(\boldsymbol{x}_{1}'\right)U^{\dagger}\left(\boldsymbol{x}_{2}'\right)U\left(\boldsymbol{x}_{3}'\right)U^{\dagger}\left(\boldsymbol{x}_{4}'\right)\right]-\frac{1}{N_{c}}{\rm tr}_{c}\left[U\left(\boldsymbol{x}_{1}'\right)U^{\dagger}\left(\boldsymbol{x}_{2}'\right)\right]{\rm tr}_{c}\left[U\left(\boldsymbol{x}_{3}'\right)U^{\dagger}\left(\boldsymbol{x}_{4}'\right)\right]\right)\right\rangle =\nonumber \\
 & \frac{1}{4}\left\langle {\rm tr}_{c}\left[U\left(\boldsymbol{x}_{1}\right)U^{\dagger}\left(\boldsymbol{x}_{2}\right)U\left(\boldsymbol{x}_{3}\right)U^{\dagger}\left(\boldsymbol{x}_{4}\right)\right]{\rm tr}_{c}\left[U\left(\boldsymbol{x}_{1}'\right)U^{\dagger}\left(\boldsymbol{x}_{2}'\right)U\left(\boldsymbol{x}_{3}'\right)U^{\dagger}\left(\boldsymbol{x}_{4}'\right)\right]\right\rangle \nonumber \\
 & -\frac{1}{4N_{c}}\left\langle {\rm tr}_{c}\left[U\left(\boldsymbol{x}_{1}\right)U^{\dagger}\left(\boldsymbol{x}_{2}\right)U\left(\boldsymbol{x}_{3}\right)U^{\dagger}\left(\boldsymbol{x}_{4}\right)\right]{\rm tr}_{c}\left[U\left(\boldsymbol{x}_{1}'\right)U^{\dagger}\left(\boldsymbol{x}_{2}'\right)\right]{\rm tr}_{c}\left[U\left(\boldsymbol{x}_{3}'\right)U^{\dagger}\left(\boldsymbol{x}_{4}'\right)\right]\right\rangle \nonumber \\
 & -\frac{1}{4N_{c}}\left\langle {\rm tr}_{c}\left[U\left(\boldsymbol{x}_{1}\right)U^{\dagger}\left(\boldsymbol{x}_{2}\right)\right]{\rm tr}_{c}\left[U\left(\boldsymbol{x}_{3}\right)U^{\dagger}\left(\boldsymbol{x}_{4}\right)\right]{\rm tr}_{c}\left[U\left(\boldsymbol{x}_{1}'\right)U^{\dagger}\left(\boldsymbol{x}_{2}'\right)U\left(\boldsymbol{x}_{3}'\right)U^{\dagger}\left(\boldsymbol{x}_{4}'\right)\right]\right\rangle \nonumber \\
 & +\left(\frac{1}{2N_{c}}\right)^{2}\left\langle {\rm tr}_{c}\left[U\left(\boldsymbol{x}_{1}\right)U^{\dagger}\left(\boldsymbol{x}_{2}\right)\right]{\rm tr}_{c}\left[U\left(\boldsymbol{x}_{3}\right)U^{\dagger}\left(\boldsymbol{x}_{4}\right)\right]{\rm tr}_{c}\left[U\left(\boldsymbol{x}_{1}'\right)U^{\dagger}\left(\boldsymbol{x}_{2}'\right)\right]{\rm tr}_{c}\left[U\left(\boldsymbol{x}_{3}'\right)U^{\dagger}\left(\boldsymbol{x}_{4}'\right)\right]\right\rangle \nonumber 
\end{align}
for the density matrix which describes interaction of the 4-quark
ensemble with the target , with subsequent formation of two color
singlet dipoles. The angular brackets $\langle...\rangle$ denote
averaging over the color sources, as defined in~(\ref{eq:Ave}).
As we demonstrate in Appendix~\ref{sec:Multipole}, after convolution
with color generators and color averaging, it is possible to rewrite~(\ref{eq:NDef})
as 
\begin{align}
 & \mathcal{N}\left(\boldsymbol{x}_{1},\,\boldsymbol{x}_{2},\,\boldsymbol{x}_{3},\,\boldsymbol{x}_{4};\,\boldsymbol{x}_{1}',\,\boldsymbol{x}_{2}',\,\boldsymbol{x}_{3}',\,\boldsymbol{x}_{4}'\right)=\label{eq:NDef-5}\\
 & =\frac{N_{c}^{2}}{4}\left\langle \left[S_{4}\left(\boldsymbol{x}_{1},\,\boldsymbol{x}_{2},\,\boldsymbol{x}_{3},\,\boldsymbol{x}_{4}\right)-S_{2}\left(\boldsymbol{x}_{1},\,\boldsymbol{x}_{2}\right)S_{2}\left(\boldsymbol{x}_{3},\,\boldsymbol{x}_{4}\right)\right]\left[S_{4}\left(\boldsymbol{x}_{1}',\,\boldsymbol{x}_{2}',\,\boldsymbol{x}_{3}',\,\boldsymbol{x}_{4}'\right)-S_{2}\left(\boldsymbol{x}_{1}',\,\boldsymbol{x}_{2}'\right)S_{2}\left(\boldsymbol{x}_{3}',\,\boldsymbol{x}_{4}'\right)\right]\right\rangle .\nonumber 
\end{align}
which includes only dipole and quadrupole correlators. In the usual
large-$N_{c}$ limit the averaging may be applied separately to different
terms in the product in~(\ref{eq:NDef-5}), so eventually the cross-section
probes the quadrupole and dipole forward scattering amplitudes. 
We need to mention that the combination of quadrupole and dipoles,
$S_{4}\left(\boldsymbol{x}_{1},\,\boldsymbol{x}_{2},\,\boldsymbol{x}_{3},\,\boldsymbol{x}_{4}\right)-S_{2}\left(\boldsymbol{x}_{1},\,\boldsymbol{x}_{2}\right)S_{2}\left(\boldsymbol{x}_{3},\,\boldsymbol{x}_{4}\right)$,
which appears bilinearly in~(\ref{eq:NDef-5}), also controls the
dominant color singlet contribution in inclusive hadroproduction of
single quarkonia~\cite{Kang:2013hta}. In the latter channel the
$p_{T}$-dependence is predominanlty sensitive to the combination
$\boldsymbol{x}_{1}+\boldsymbol{x}_{2}-\boldsymbol{x}_{3}-\boldsymbol{x}_{4}$
(difference of impact parameters of the produced quarkonia in the
amplitude and its conjugate), whereas dependence on other combinations
of transverse coordinates is integrated out in the heavy quark mass
limit~\footnote{In our qualitative discussion of single quarkonia production we disregard
additional phase factors $\sim e^{\pm i\boldsymbol{p}\cdot\boldsymbol{r}_{ij}}$
which take into account transverse boosts of the quarkonium wave function~\cite{Kowalski:2003hm,Kowalski:2006hc}.
For quarkonia the typical dipole size $\boldsymbol{r}_{ij}$ is controlled
by the heavy quark mass and becomes negligible in the heavy quark
mass limit.}. The double quarkonia production cross-section~(\ref{eq:NDef-5})
includes convolution with completely different dependence on coordinates,
and for this reason provides independent observable for study of the
quadrupole contributions. In the special limit when the distance between
any quark-antiquark pair vanishes, the quadrupole amplitude reduces
to a dipole amplitude, and for this reason~(\ref{eq:NDef-5}) cancels
exactly. Due to this, the integral~(\ref{eq:XSec}) remains finite
despite of possible singular behavior of the wave function $\psi_{\bar{Q}Q\bar{Q}Q}^{(\gamma)}$
in this limit.

While the differential cross-section~(\ref{eq:XSec}) contains exhaustive
information about the multipole elements of the target, due to smallness
of the cross-sections, experimentally it may be interesting to study
the $\boldsymbol{p}_{\perp}$-integrated cross-section, for which
analytic integration over $\int d^{2}\boldsymbol{p}_{1}^{\perp}$
and $\int d^{2}\boldsymbol{p}_{2}^{\perp}$ yields~
\begin{align}
\frac{d\sigma}{dy_{1}\,dy_{2}}= & 4\left(2\pi\right)^{3}\int d^{2}\boldsymbol{\xi}\,d^{2}\boldsymbol{\eta}\left(\prod_{i=1}^{4}d^{2}\boldsymbol{x}_{i}\right)\mathcal{N}\left(\boldsymbol{x}_{1},\,\boldsymbol{x}_{2},\,\boldsymbol{x}_{3},\,\boldsymbol{x}_{4};\,\boldsymbol{\xi},\,\boldsymbol{x}_{1}+\boldsymbol{x}_{2}-\boldsymbol{\xi},\,\boldsymbol{\eta},\,\boldsymbol{x}_{3}+\boldsymbol{x}_{4}-\boldsymbol{\eta}\right)\times\label{eq:XSec-5}\\
\times & \left[\Psi_{M_{1}}^{\dagger h_{1}h_{2}}\Psi_{M_{2}}^{\dagger h_{3}h_{4}}\psi_{\bar{Q}Q\bar{Q}Q}^{(\gamma)h_{1}h_{2}h_{3}h_{4}}\left(\boldsymbol{x}_{1};\,\boldsymbol{x}_{2};\,\boldsymbol{x}_{3};\,\boldsymbol{x}_{4};\,q\right)\right]\times\nonumber \\
\times & \left(\Psi_{M_{1}}^{\dagger\mathfrak{h}_{1}\mathfrak{h}_{2}}\Psi_{M_{2}}^{\dagger\mathfrak{h}_{3}\mathfrak{h}_{4}}\psi_{\bar{Q}Q\bar{Q}Q}^{(\gamma)\mathfrak{h}_{1}\mathfrak{h}_{2}\mathfrak{h}_{3}\mathfrak{h}_{4}}\left(\boldsymbol{\xi};\,\boldsymbol{x}_{1}+\boldsymbol{x}_{2}-\boldsymbol{\xi};\,\boldsymbol{\eta};\,\boldsymbol{x}_{3}+\boldsymbol{x}_{4}-\boldsymbol{\eta};\,q\right)\right)^{*}.\nonumber
\end{align}

Finally, we would like to discuss briefly evaluation of the diagrams
shown in the Figure~\ref{fig:QuarkoniaPairCGC-1}, which can contribute
in general case of quarkonia pair production. The evaluation largely
follows the same procedure, however the interaction of the shock wave
with heavy quarks at early stages (diagrams in the left and middle
columns of the Figure~\ref{fig:QuarkoniaPairCGC-1}) does not allow
to rewrite the cross-section as a convolution of $\psi_{\bar{Q}Q\bar{Q}Q}^{(\gamma)}$
with universal target-dependent structure even in eikonal approximation:
due to peculiar structure of the shock wave interaction with partons~(\ref{eq:SWubar}-\ref{SWg}),
there is no contributions with instantaneous virtual quarks and gluons
which were taken into account in evaluation of~$\psi_{\bar{Q}Q\bar{Q}Q}^{(\gamma)}$.
Since in the final state we register only the color singlet heavy
quarkonia ($\approx$ take traces over the color indices of heavy
quark lines, independently in the amplitude and its conjugate), the
interaction with the target eventually can be expressed in terms of
the dipole and quadrupole amplitudes (no sextupole and octupole contributions),
however the structure of such expressions will be different for diagrams
in each column of the Figure~\ref{fig:QuarkoniaPairCGC-1}. A systematic
evaluations for that case requires a dedicated study and will be presented
elsewhere.

\section{Phenomenological estimates}

\label{sec:Numer}

\subsection{Parametrization of the dipole and quadrupole amplitudes}

\label{sec:Parametrization-S4}The numerical estimates depend crucially
on the choice of the parametrization of the dipole and quadrupole
amplitudes. For the dipole amplitude $N$, we will use the phenomenological
impact parameter $b$-dependent ``bCGC'' parametrization~\cite{Kowalski:2006hc,RESH}
which has correct asymptotic behavior in the small- and large-size
dipole limits, and has been fitted to reproduce various phenomenological
data with reasonable precision.

The quadrupole amplitude in general is an independent nonperturbative
object, not related to a dipole amplitude. However, as was discussed
in~\cite{Iancu:2011nj,Iancu:2011ns,Iancu:2012xa}, in the limit
of large number of colors $N_{c}\gg1$, it is possible to relate various
multipole and multitrace matrix elements, and essentially express
them all in terms of the dipole amplitude $N$. For heavy charmonia
in QCD, the use of the large-$N_{c}$ limit is well-justified, since
the discarded $\sim\mathcal{O}\left(1/N_{c}\right)$ corrections parametrically
are on par with strong coupling $\alpha_{s}\left(m_{c}\right)\sim1/3$.
The derivation of these identities is cumbersome and may be found
in the above-mentioned references~\cite{Iancu:2011nj,Iancu:2011ns}.

The possibility to express the amplitude in terms of the color singlet
dipoles may be also understood from the \textit{area enhancement
argument} introduced in~\cite{Agostini:2021xc,Kovner:2017ssr,Kovner:2018vec}:
The interaction of the color singlet multipole (e.g. quadrupole) with
the target grows as a function of its size due to color transparency,
for this reason we expect that in the phase space integration, the
dominant contribution should come from the configurations with well-separated
partons in the transverse space. On the other hand, color group averaging
suppresses correlators in which the distance between partons exceeding
the color correlation length set by saturation scale, namely $\left|\boldsymbol{x}_{i}-\boldsymbol{x}_{j}\right|\gtrsim Q_{s}^{-1}$.
However, this suppression does not work for configurations in which
all partons are grouped into small-size color singlet pairs, separated
by large distances between pairs~\footnote{Formally in our expansion we should take into account all groups of
matrices which include color singlet irreducible representation in
expansion of direct product, for example color triplets $\left\langle U\left(\boldsymbol{x}_{1}\right)U\left(\boldsymbol{x}_{2}\right)U\left(\boldsymbol{x}_{3}\right)\right\rangle $
for $N_{c}=3$. However, from area enhancement argument it is clear
that the dominant contribution comes from the cofigurations with maximal
number of disjoint blocks, which corresponds to pairwise products
$U^{\dagger}\left(\boldsymbol{x}_{1}\right)U\left(\boldsymbol{x}_{2}\right)$.}. After the integration over phase space, it becomes clear that such
multi-pair configurations are enhanced compared to true or ``genuine''
multipoles (quadrupoles). The enhancement factor is given by $\sim\left(S_{\perp}Q_{s}^{2}\right)^{n-1}$,
where $S_{\perp}$ is the transverse area of the target, and $n\gtrsim1$
is the number of color singlet pairs. According to modern phenomenological
parametrizations, the typical values of saturation scale $Q_{s}$
in the kinematics of interest are of order 1 GeV$\sim\left(0.2\,{\rm fm}\right)^{-1}$,
for this reason the enhancement factor $\left(S_{\perp}Q_{s}^{2}\right)$
might be numerically large even for the proton. For this reason, any
multipoint correlator may be approximated as a sum of pairwise products~\cite{Agostini:2021xc,Kovner:2017ssr,Kovner:2018vec}
\begin{align}
\left\langle U^{\dagger}\left(\boldsymbol{y}_{1}\right)^{a_{1}b_{1}}U\left(\boldsymbol{y}_{2}\right)^{a_{2}b_{2}}...U^{\dagger}\left(\boldsymbol{y}_{2n-1}\right)^{a_{2n-1}b_{2n-1}}U\left(\boldsymbol{y}_{2n}\right)^{a_{2n}b_{2n}}\right\rangle  & \approx\sum_{\sigma\in\Pi(\chi)}\prod_{\{\alpha,\beta\}\in\sigma}\left\langle U^{\dagger}\left(\boldsymbol{y}_{\alpha}\right)^{a_{\alpha}b_{\alpha}}U\left(\boldsymbol{y}_{\beta}\right)^{a_{\beta}b_{\beta}}\right\rangle ,\label{eq:Wick}
\end{align}
where $\chi=\{1,2,...,2n\}$ and $\Pi(\chi)$ is the set of partitions
of $\chi$ with disjoint pairs. The result~(\ref{eq:Wick}) significantly
simplifies evaluation of the diagrams with multipole contributions
and allows to apply Wick's theorem in order to introduce glasma graph
approach. 

As was suggested in~\cite{Iancu:2011nj,Iancu:2011ns}, the result~(\ref{eq:Wick})
for the the quadrupoles might be replaced with more accurate expression
which takes into account the effects of JIMWLK evolution, as well
as has more accurate behavior in a few physically important limits.
Explicitly, the amplitude $\overline{S_{4}}$ is given by 
\begin{align}
\overline{S_{4}} & \equiv\left\langle \frac{1}{N_{c}}{\rm tr}_{c}\left[U\left(\boldsymbol{x}_{1}\right)U^{\dagger}\left(\boldsymbol{x}_{2}\right)U\left(\boldsymbol{x}_{3}\right)U^{\dagger}\left(\boldsymbol{x}_{4}\right)\right]\right\rangle \approx\label{eq:S4Param}\\
 & \approx\frac{\Gamma_{Y}\left(\boldsymbol{x}_{1},\,\boldsymbol{x}_{2}\right)+\Gamma_{Y}\left(\boldsymbol{x}_{3},\,\boldsymbol{x}_{4}\right)-\Gamma_{Y}\left(\boldsymbol{x}_{1},\,\boldsymbol{x}_{3}\right)-\Gamma_{Y}\left(\boldsymbol{x}_{2},\,\boldsymbol{x}_{4}\right)}{\Gamma_{Y}\left(\boldsymbol{x}_{1},\,\boldsymbol{x}_{2}\right)+\Gamma_{Y}\left(\boldsymbol{x}_{3},\,\boldsymbol{x}_{4}\right)-\Gamma_{Y}\left(\boldsymbol{x}_{1},\,\boldsymbol{x}_{4}\right)-\Gamma_{Y}\left(\boldsymbol{x}_{2},\,\boldsymbol{x}_{3}\right)}\overline{S_{2}}\left(\boldsymbol{x}_{1},\,\boldsymbol{x}_{2}\right)\overline{S_{2}}\left(\boldsymbol{x}_{3},\,\boldsymbol{x}_{4}\right)+\nonumber \\
 & +\frac{\Gamma_{Y}\left(\boldsymbol{x}_{1},\,\boldsymbol{x}_{4}\right)+\Gamma_{Y}\left(\boldsymbol{x}_{2},\,\boldsymbol{x}_{3}\right)-\Gamma_{Y}\left(\boldsymbol{x}_{1},\,\boldsymbol{x}_{3}\right)-\Gamma_{Y}\left(\boldsymbol{x}_{2},\,\boldsymbol{x}_{4}\right)}{\Gamma_{Y}\left(\boldsymbol{x}_{1},\,\boldsymbol{x}_{4}\right)+\Gamma_{Y}\left(\boldsymbol{x}_{2},\,\boldsymbol{x}_{3}\right)-\Gamma_{Y}\left(\boldsymbol{x}_{1},\,\boldsymbol{x}_{2}\right)-\Gamma_{Y}\left(\boldsymbol{x}_{3},\,\boldsymbol{x}_{4}\right)}\overline{S_{2}}\left(\boldsymbol{x}_{1},\,\boldsymbol{x}_{4}\right)\overline{S_{2}}\left(\boldsymbol{x}_{2},\,\boldsymbol{x}_{3}\right)\nonumber 
\end{align}
where the function $\Gamma_{Y}\left(\boldsymbol{x}_{i},\,\boldsymbol{x}_{j}\right)$
is defined as
\begin{equation}
\Gamma_{Y}\left(\boldsymbol{x}_{i},\,\boldsymbol{x}_{j}\right)=-\ln\left(\overline{S_{2}}\left(\boldsymbol{x}_{i},\,\boldsymbol{x}_{j}\right)\right)=-\ln\left(1-\,N\left(x,\,\boldsymbol{r}_{ij},\,\boldsymbol{b}_{ij}\right)\right).
\end{equation}
The expression~(\ref{eq:S4Param}) includes denominators vanishing
at the hypersurface 
\begin{align}
\Gamma_{Y}\left(\boldsymbol{x}_{1},\,\boldsymbol{x}_{2}\right)+\Gamma_{Y}\left(\boldsymbol{x}_{3},\,\boldsymbol{x}_{4}\right) & =\Gamma_{Y}\left(\boldsymbol{x}_{1},\,\boldsymbol{x}_{4}\right)+\Gamma_{Y}\left(\boldsymbol{x}_{2},\,\boldsymbol{x}_{3}\right),\label{eq:hypersurface}
\end{align}
however the function $\overline{S}_{4}$ remains finite when approaching
this hypersurface, with limiting value given by
\begin{align}
\left.\overline{S_{4}}\right|_{(\ref{eq:hypersurface})} & \approx\overline{S_{2}}\left(\boldsymbol{x}_{1},\,\boldsymbol{x}_{2}\right)\overline{S_{2}}\left(\boldsymbol{x}_{3},\,\boldsymbol{x}_{4}\right)\left(1+\ln\left(\frac{\overline{S_{2}}\left(\boldsymbol{x}_{1},\,\boldsymbol{x}_{2}\right)\overline{S_{2}}\left(\boldsymbol{x}_{3},\,\boldsymbol{x}_{4}\right)}{\overline{S_{2}}\left(\boldsymbol{x}_{1},\,\boldsymbol{x}_{3}\right)\overline{S_{2}}\left(\boldsymbol{x}_{2},\,\boldsymbol{x}_{4}\right)}\right)\right).\label{eq:S4Param-1}
\end{align}

The combination~$\overline{S_{4}}\left(\boldsymbol{x}_{1},\,\boldsymbol{x}_{2},\,\boldsymbol{x}_{3},\,\boldsymbol{x}_{4}\right)-\overline{S_{2}}\left(\boldsymbol{x}_{1},\,\boldsymbol{x}_{2}\right)\overline{S_{2}}\left(\boldsymbol{x}_{3},\,\boldsymbol{x}_{4}\right)$
which appears in~(\ref{eq:NDef-5}) may be rewritten as
\begin{align}
\overline{S_{4}} & \left(\boldsymbol{x}_{1},\,\boldsymbol{x}_{2},\,\boldsymbol{x}_{3},\,\boldsymbol{x}_{4}\right)-\overline{S_{2}}\left(\boldsymbol{x}_{1},\,\boldsymbol{x}_{2}\right)\overline{S_{2}}\left(\boldsymbol{x}_{3},\,\boldsymbol{x}_{4}\right)\approx\label{eq:S4Param-2}\\
 & =\frac{\Gamma_{Y}\left(\boldsymbol{x}_{1},\,\boldsymbol{x}_{4}\right)+\Gamma_{Y}\left(\boldsymbol{x}_{2},\,\boldsymbol{x}_{3}\right)-\Gamma_{Y}\left(\boldsymbol{x}_{1},\,\boldsymbol{x}_{3}\right)-\Gamma_{Y}\left(\boldsymbol{x}_{2},\,\boldsymbol{x}_{4}\right)}{\Gamma_{Y}\left(\boldsymbol{x}_{1},\,\boldsymbol{x}_{4}\right)+\Gamma_{Y}\left(\boldsymbol{x}_{2},\,\boldsymbol{x}_{3}\right)-\Gamma_{Y}\left(\boldsymbol{x}_{1},\,\boldsymbol{x}_{2}\right)-\Gamma_{Y}\left(\boldsymbol{x}_{3},\,\boldsymbol{x}_{4}\right)}\left[\overline{S_{2}}\left(\boldsymbol{x}_{1},\,\boldsymbol{x}_{2}\right)\overline{S_{2}}\left(\boldsymbol{x}_{3},\,\boldsymbol{x}_{4}\right)+\overline{S_{2}}\left(\boldsymbol{x}_{1},\,\boldsymbol{x}_{4}\right)\overline{S_{2}}\left(\boldsymbol{x}_{2},\,\boldsymbol{x}_{3}\right)\right]\nonumber 
\end{align}
In the dilute limit, the expression for $\overline{S_{4}}$ simplifies
as~\cite{Iancu:2011nj,Iancu:2011ns}

\begin{align}
\overline{S_{4}} & \equiv\left\langle \frac{1}{N_{c}}{\rm tr}_{c}\left[U\left(\boldsymbol{x}_{1}\right)U^{\dagger}\left(\boldsymbol{x}_{2}\right)U\left(\boldsymbol{x}_{3}\right)U^{\dagger}\left(\boldsymbol{x}_{4}\right)\right]\right\rangle \approx1-\,N\left(x,\,\boldsymbol{r}_{12},\,\boldsymbol{b}_{12}\right)-\,N\left(x,\,\boldsymbol{r}_{34},\,\boldsymbol{b}_{34}\right)\label{eq:S4dilute}\\
 & -N\left(x,\,\boldsymbol{r}_{23},\,\boldsymbol{b}_{23}\right)-N\left(x,\,\boldsymbol{r}_{14},\,\boldsymbol{b}_{14}\right)+N\left(x,\,\boldsymbol{r}_{13},\,\boldsymbol{b}_{13}\right)+N\left(x,\,\boldsymbol{r}_{24},\,\boldsymbol{b}_{24}\right).\nonumber 
\end{align}
Substituting this result into the amplitude~(\ref{eq:NDef}), it
is possible to show that the nonzero contributions show up only when
we take into account the nonlinear $\sim\mathcal{O}\left(N^{2}\right)$
terms in expansion and reads as 

\begin{align}
 & \mathcal{N}\left(\boldsymbol{x}_{1},\,\boldsymbol{x}_{2},\,\boldsymbol{x}_{3},\,\boldsymbol{x}_{4};\,\boldsymbol{x}_{1}',\,\boldsymbol{x}_{2}',\,\boldsymbol{x}_{3}',\,\boldsymbol{x}_{4}'\right)\approx\label{eq:NDef-3}\\
 & \approx\frac{N_{c}^{2}}{4}\left[N\left(x,\,\boldsymbol{r}_{13},\,\boldsymbol{b}_{13}\right)+N\left(x,\,\boldsymbol{r}_{24},\,\boldsymbol{b}_{24}\right)-N\left(x,\,\boldsymbol{r}_{23},\,\boldsymbol{b}_{23}\right)-N\left(x,\,\boldsymbol{r}_{14},\,\boldsymbol{b}_{14}\right)\right]\times\nonumber \\
 & \times\left[N\left(x,\,\boldsymbol{r}_{13}',\,\boldsymbol{b}_{13}'\right)+N\left(x,\,\boldsymbol{r}_{24}',\,\boldsymbol{b}_{24}'\right)-N\left(x,\,\boldsymbol{r}_{23}',\,\boldsymbol{b}_{23}'\right)-N\left(x,\,\boldsymbol{r}_{14}',\,\boldsymbol{b}_{14}'\right)\right].\nonumber 
\end{align}

The expressions in the second and the third lines of~(\ref{eq:NDef-3})
may be rewritten using identity
\begin{align}
 & N\left(x,\,\boldsymbol{r}_{13},\,\boldsymbol{b}_{13}\right)+N\left(x,\,\boldsymbol{r}_{24},\,\boldsymbol{b}_{24}\right)-N\left(x,\,\boldsymbol{r}_{23},\,\boldsymbol{b}_{23}\right)-N\left(x,\,\boldsymbol{r}_{14},\,\boldsymbol{b}_{14}\right)\approx\label{eq:NDef-4-1}\\
 & \approx\left[N\left(x,\,\boldsymbol{r}_{13},\,\boldsymbol{b}_{13}\right)-N\left(x,\,\boldsymbol{r}_{23},\,\boldsymbol{b}_{23}\right)\right]-\left[N\left(x,\,\boldsymbol{r}_{14},\,\boldsymbol{b}_{14}\right)-N\left(x,\,\boldsymbol{r}_{24},\,\boldsymbol{b}_{24}\right)\right]\nonumber 
\end{align}
which clearly shows that~(\ref{eq:NDef-3}) vanishes when any of
the dipole sizes $\boldsymbol{r}_{12},\,\boldsymbol{r}_{34},\,\boldsymbol{r}_{12}',\,\boldsymbol{r}_{34}'$
goes to zero. Indeed, as was discussed in previous section, in this
limit the quadrupole amplitude reduces to a dipole amplitude and exactly
cancels in~(\ref{eq:NDef-5}). In the small-$r$ domain it is expected
that the dipole amplitude $N$ should have an asymptotic behavior
\begin{equation}
N\left(x,\,\boldsymbol{r},\,\boldsymbol{b}\right)\approx\left(Q_{s}^{2}\left(x,\,\boldsymbol{b}\right)r^{2}\right)^{\gamma},
\end{equation}
where $Q_{s}$ is the saturation scale, and $\gamma\lesssim1$ is
some numerical coefficient. For $\gamma\approx1$ (\textit{e.g}.~``GBW''
and ``bSat'' parametrizaitons~\cite{Rezaeian:2012ji}) it is possible
to simplify drastically~(\ref{eq:NDef-3}) and rewrite it as 
\begin{align}
 & \mathcal{N}^{(\gamma=1)}\left(\boldsymbol{x}_{1},\,\boldsymbol{x}_{2},\,\boldsymbol{x}_{3},\,\boldsymbol{x}_{4};\,\boldsymbol{x}_{1}',\,\boldsymbol{x}_{2}',\,\boldsymbol{x}_{3}',\,\boldsymbol{x}_{4}'\right)\approx\frac{N_{c}^{2}Q_{s}^{4}\left(x\right)}{2}\left(\boldsymbol{r}_{12}\cdot\boldsymbol{r}_{34}\right)\left(\boldsymbol{r}_{12}'\cdot\boldsymbol{r}_{34}'\right),\label{eq:NDef-4}
\end{align}
which demonstrates the dependence of the scattering amplitude $\mathcal{N}$
on sizes of individual dipoles in heavy quark mass limit. For $\gamma\not=1$
a simple form~(\ref{eq:NDef-4}) is not valid. However the antisymmetry
of~(\ref{eq:NDef-4-1}) with respect to permutations $\boldsymbol{r}_{1}\leftrightarrow\boldsymbol{r}_{2},\,\boldsymbol{r}_{3}\leftrightarrow\boldsymbol{r}_{4}$
indicates that the amplitude~(\ref{eq:NDef-3}) should have a pronounced
sensitivity to a relative orientation of vectors $\left(\boldsymbol{r}_{12},\,\boldsymbol{r}_{34}\right)$
and $\left(\boldsymbol{r}_{12}',\,\boldsymbol{r}_{34}'\right)$. 

\subsection{Numerical estimates for cross-sections}

For the sake of definiteness, we'll focus on the production of the
$J/\psi+\eta_{b}$ and $\Upsilon(1S)+\eta_{c}$ production, which
are expected to have the largest cross-section among all the charmonia-bottomonia
pairs. We disregard the color octet channels in view of the smallness
of color octet LDMEs (see however discussion in~\cite{Butenschoen:2014dra,Feng:2015cba,Baranov:2015laa,Baranov:2016clx}
about their significant uncertainty). For the dominant color singlet
LDMEs we will use the values~\cite{Feng:2015wka,Abdulov:2020nxh,Baranov:2016clx}
\begin{align}
\left\langle \mathcal{O}^{J/\psi}\left(^{3}S_{1}^{[1]}\right)\right\rangle \approx3\left\langle \mathcal{O}^{\eta_{c}}\left(^{1}S_{0}^{[1]}\right)\right\rangle  & \approx1.16\,{\rm GeV}^{3},\label{eq:CO1-2}\\
\left\langle \mathcal{O}^{\Upsilon(1S)}\left(^{3}S_{1}^{[1]}\right)\right\rangle \approx3\left\langle \mathcal{O}^{\eta_{b}}\left(^{1}S_{0}^{[1]}\right)\right\rangle  & \approx8.39\,{\rm GeV}^{3},
\end{align}
where the LDMEs of pseudoscalar mesons are fixed using the relations
valid in heavy quark mass limit~\cite{Bodwin:1994jh}. As expected
from potential models~\cite{Eichten:1978tg,Eichten:1979ms,Quigg:1977dd,Martin:1980jx},
these LDMEs approximately scale with heavy quark mass as $\sim\left(\alpha_{s}\left(m_{Q}\right)m_{Q}\right)^{3}$.

\begin{figure}
\includegraphics[width=9cm]{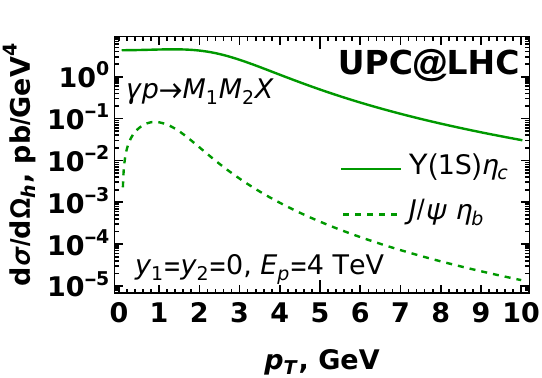}\includegraphics[width=9cm]{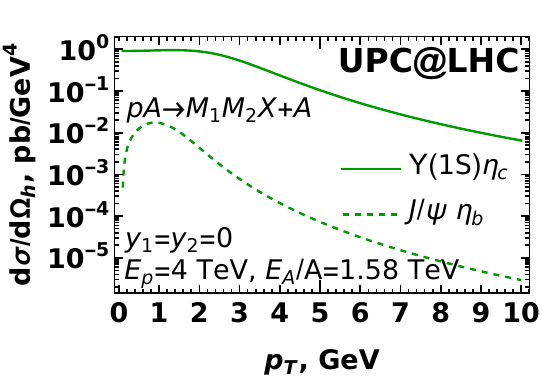}

\includegraphics[width=9cm]{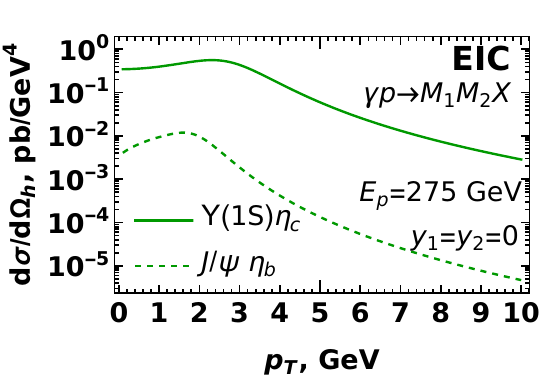}\includegraphics[width=9cm]{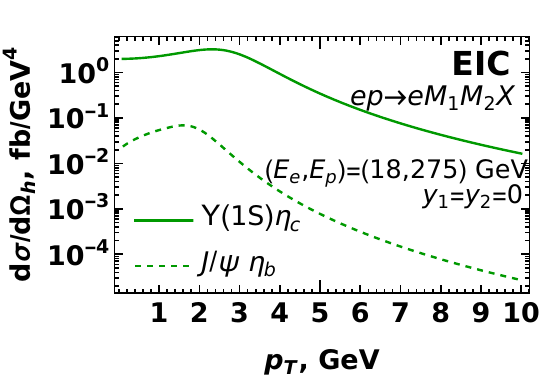}

\caption{\label{fig:RelRoleJP}The cross-sections of inclusive production of
different quarkonia pairs with opposite $C$-parities. For the sake
of definiteness we considered that both quarkonia are produced with
the same absolute value of the transverse momenta $p_{1}^{\perp}=p_{2}^{\perp}=p_{T}$.
Upper and lower rows correspond to kinematics of ultraperipheral collisions and the Electron-Ion Collider, respectively. The left column corresponds to cross-sections
of the photon-proton subprocess, the right column includes predictions for the cross-section of the full process. For ultraperipheral collisions we assume proton-lead collisions with $A=82$. For EIC we considered the electron-proton collisions only; for heavy nuclei these results should be understood as cross-sections per nucleon. For the sake of definiteness we considered
production at central rapidities ($y_{1}=y_{2}=0$) in the lab frame.
The shape of the $p_{T}$-dependence has a very mild dependence on
rapidity and azimuthal angle $\phi$ between the produced quarkonia.}
\end{figure}

\begin{figure}
\includegraphics[height=6cm]{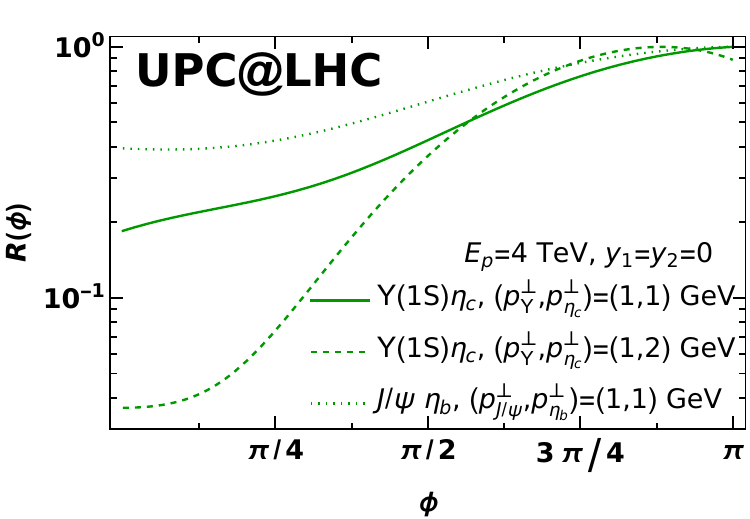}\includegraphics[height=6cm]{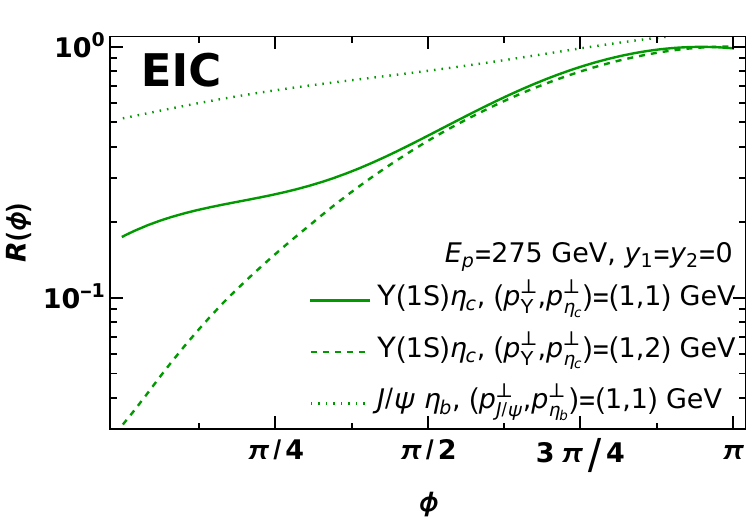}

\includegraphics[height=6cm]{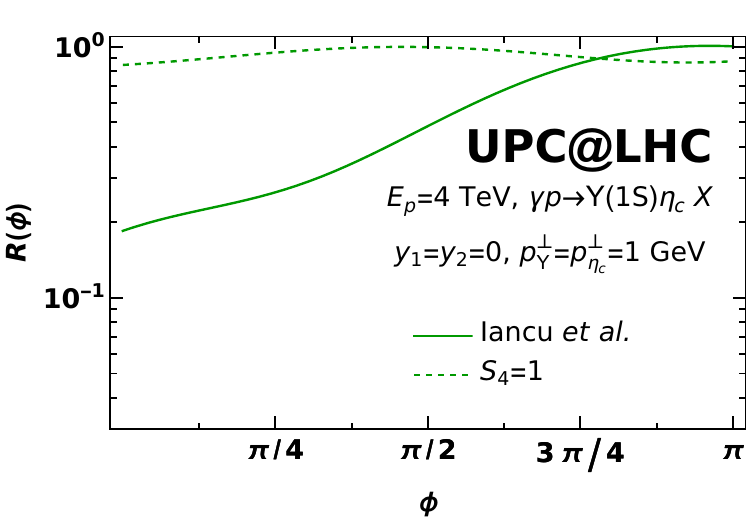}\includegraphics[height=6cm]{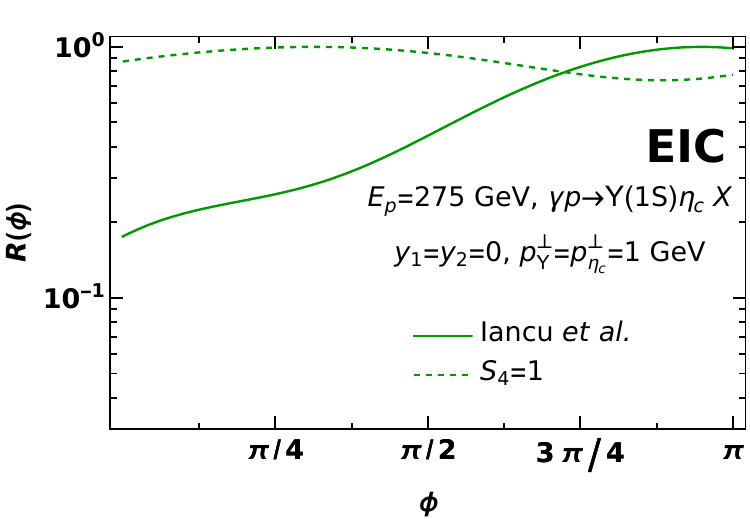}

\caption{\label{fig:RelRolePhi}Upper row: Dependence of the normalized ratio
$R(\phi)$, defined in~(\ref{eq:RatioR}), on the angle $\phi$ (azimuthal
angle between transverse momenta of quarkonia). The left plot corresponds
to the kinematics of ultraperipheral collisions at LHC, the right
plot is for the kinematics of the Electron Ion Collider. The cross-section
has almost the same angular dependence for all rapidities and transverse
momenta $p_{T}$ (see the text for more detailed explanation). Lower
row: Comparison of angular dependence found with parametrization (\ref{eq:S4Param})
(curve with label \textquotedblleft Iancu \textit{et al}.\textquotedblright )
and a trivial parametrization~(\ref{eq:S4Tri}).}
\end{figure}

In Figure~\ref{fig:RelRoleJP} we illustrate the transverse momentum
dependence of the cross-section for different quarkonia states. The
significant difference of $J/\psi\,\eta_{b}$ and $\Upsilon(1S)\,\eta_{c}$
cross-sections might be understood in the dilute limit: in this case
the pseudoscalar quarkonium can originate only from the secondary
(lower) quark loop in the Figure~\ref{fig:QuarkoniaPairCGC}. The
virtuality of the gluon which connects the two loops is largely controlled
by the mass of this pseudoscalar quarkonium, thus yielding a relative
suppression factor
\begin{equation}
\frac{d\sigma_{\gamma p\to J/\psi\,\eta_{b}}}{d\sigma_{\gamma p\to\Upsilon(1S)\eta_{c}}}\sim\left(\frac{M_{\eta_{c}}}{M_{\eta_{b}}}\right)^{4}\sim10^{-2}.
\end{equation}
 In Figure~\ref{fig:RelRolePhi} we study the dependence of the cross-sections
on the azimuthal angle $\phi$ between the transverse momenta of the
two quarkonia states. In order to make meaningful comparison of the
cross-sections, which differ by orders of magnitude, we plotted the
normalized ratio of the cross-sections,
\begin{align}
R(\phi) & =\frac{d\sigma\left(...,\,\phi\right)/d\Omega_{h}}{d\sigma\left(...,\,\phi_{{\rm max}}\right)/d\Omega_{h}},\label{eq:RatioR}
\end{align}
where $\phi_{{\rm max}}$ is the angle which maximizes the numerators
of~(\ref{eq:RatioR}). Since the quarkonia (dipole) sizes are small
in the heavy quark mass limit, the dependence on transverse momenta
in the small-$p_{T}$ kinematics is largely sensitive to the dependence
on the (Fourier conjugate) impact parameters $\boldsymbol{b}_{12}=(\boldsymbol{x}_{1}+\boldsymbol{x}_{2})/2$
and $\boldsymbol{b}_{34}=(\boldsymbol{x}_{3}+\boldsymbol{x}_{4})/2$
in the implemented quadrupole amplitude. For the parametrization~(\ref{eq:S4Param}),
the dependence on angle $\varphi$ between the centers of mass of
the dipoles, $\boldsymbol{b}_{12}$ and $\boldsymbol{b}_{34}$, is
largely controlled by prefactor

\begin{align}
 & \sim\Gamma_{Y}\left(\boldsymbol{x}_{1},\,\boldsymbol{x}_{4}\right)+\Gamma_{Y}\left(\boldsymbol{x}_{2},\,\boldsymbol{x}_{3}\right)-\Gamma_{Y}\left(\boldsymbol{x}_{1},\,\boldsymbol{x}_{3}\right)-\Gamma_{Y}\left(\boldsymbol{x}_{2},\,\boldsymbol{x}_{4}\right)\approx\Gamma_{Y}\left(\boldsymbol{b}_{12}-\boldsymbol{b}_{34}+\frac{\boldsymbol{r}_{12}+\boldsymbol{r}_{34}}{2}\right)+\label{eq:Fact}\\
 & +\Gamma_{Y}\left(\boldsymbol{b}_{12}-\boldsymbol{b}_{34}-\frac{\boldsymbol{r}_{12}+\boldsymbol{r}_{34}}{2}\right)-\Gamma_{Y}\left(\boldsymbol{b}_{12}-\boldsymbol{b}_{34}+\frac{\boldsymbol{r}_{12}-\boldsymbol{r}_{34}}{2}\right)-\Gamma_{Y}\left(\boldsymbol{b}_{12}-\boldsymbol{b}_{34}-\frac{\boldsymbol{r}_{12}-\boldsymbol{r}_{34}}{2}\right),\nonumber 
\end{align}
where we introduced variables $\boldsymbol{r}_{ij}=\boldsymbol{x}_{i}-\boldsymbol{x}_{j}$
for the distance between pairs of the heavy quarks. In the heavy quark
mass limit, the dominant contribution comes from the region $r_{12},r_{34}\ll b_{12},b_{34}$.
For the angle $\varphi\approx0$ (collinearly directed impact parameters),
the factor~(\ref{eq:Fact}) is suppressed at $\boldsymbol{b}_{12}\approx\boldsymbol{b}_{34}$:
physically this corresponds to a tiny quadrupole passing at some distance
from the nucleus. The cross-section increases homogeneously as a function
of $\phi$, which merely reflects growth of the quadrupole moment
of the four-quark ensemble. The predicted pronounced $\phi$-dependence
suggests that predominantly the production occurs in the back-to-back
kinematics, akin to exclusive photoproduction of the same pairs~\cite{Andrade:2022wph}.
 In order to demonstrate that the expected $\phi$-dependence is
due to the implemented parametrization of the quadrupole amplitude,
in the second row of the Figure~\ref{fig:RelRolePhi} we compare
predictions of the quadrupole parametrization (\ref{eq:S4Param})
with predictions obtained with a trivial parametrization 
\begin{equation}
S_{4}\left(\boldsymbol{x}_{1},\boldsymbol{x}_{2},\boldsymbol{x}_{3},\boldsymbol{x}_{4}\right)=1,\label{eq:S4Tri}
\end{equation}
which corresponds to a free noninteracting quadrupole. For the latter
parametrization, integration of the quadrupole terms in~(\ref{eq:XSec})
yields $\delta$-functions $\sim\delta\left(\boldsymbol{p}_{1}^{\perp}\right),\,\delta\left(\boldsymbol{p}_{2}^{\perp}\right)$,
so for nonzero momenta $\boldsymbol{p}_{1}^{\perp},\boldsymbol{p}_{2}^{\perp}$
the quadrupole contribution effectively drops out. As we can see from
the last row in the Figure~\ref{fig:RelRolePhi}, the $\phi$-dependence in this case is negligibly small:
it exists due to a mild dependence on orientation of the two dipoles
$\boldsymbol{r}_{12},\,\boldsymbol{r}_{34}$ in the wave function
$\psi_{\bar{Q}Q\bar{Q}Q}^{(\gamma)}$. This finding corroborates that
the predicted $\phi$-dependence is due to the chosen parametrization
of the quadrupole amplitude.

\begin{figure}
\includegraphics[width=9cm]{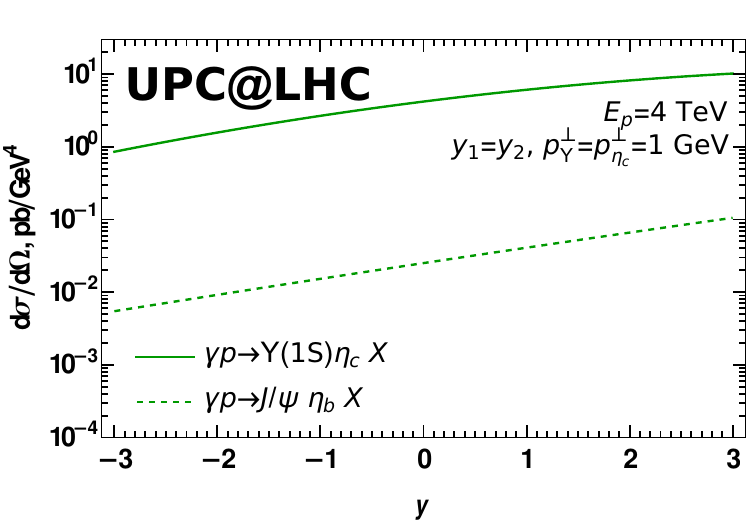}\includegraphics[width=9cm]{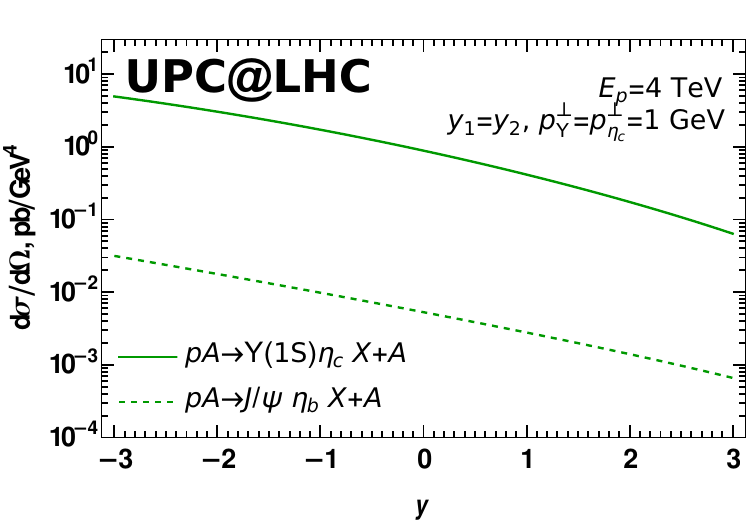}

\includegraphics[width=9cm]{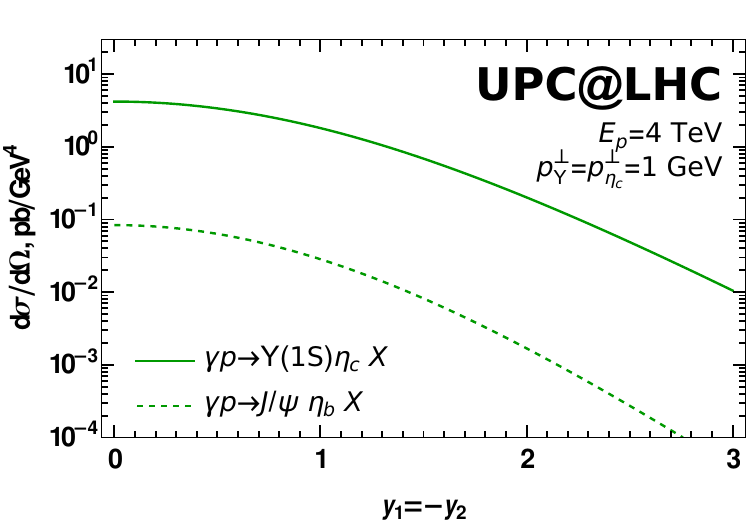}\includegraphics[width=9cm]{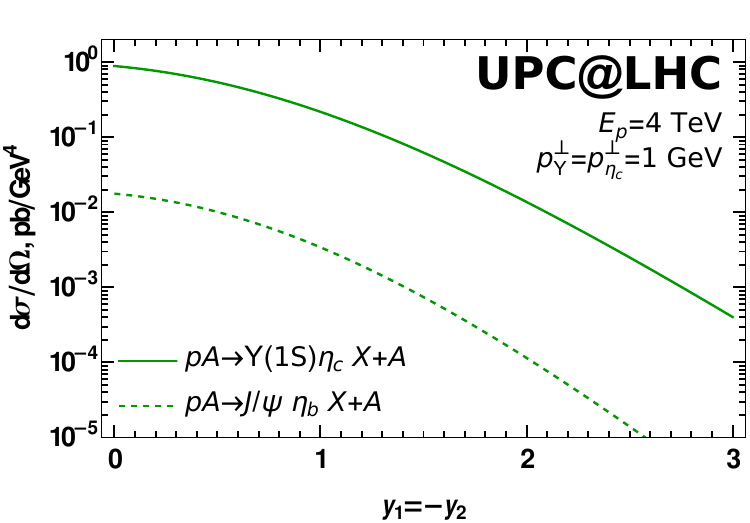}

\caption{\label{fig:Rapidities-UPC}The rapidity dependence of the cross-section
in the kinematics of the ultraperipheral collisions at LHC. The plots
in the upper row correspond to a configuration with equal rapidities
of the produced quarkonia, $y_{1}=y_{2}$, whereas the lower row corresponds
to rapidities which differ by a sign in lab frame, $y_{1}=-y_{2}$.
In both rows the left plot corresponds to the cross-section of the
\textit{photo}production subprocess, and the right column shows predictions
for the cross-section of the full process $pA\to A+M_{1}M_{2}X$,
assuming proton-lead collisions, as defined in~(\ref{eq:EPA_q-2-1}).}
\end{figure}

\begin{figure}
\includegraphics[width=9cm]{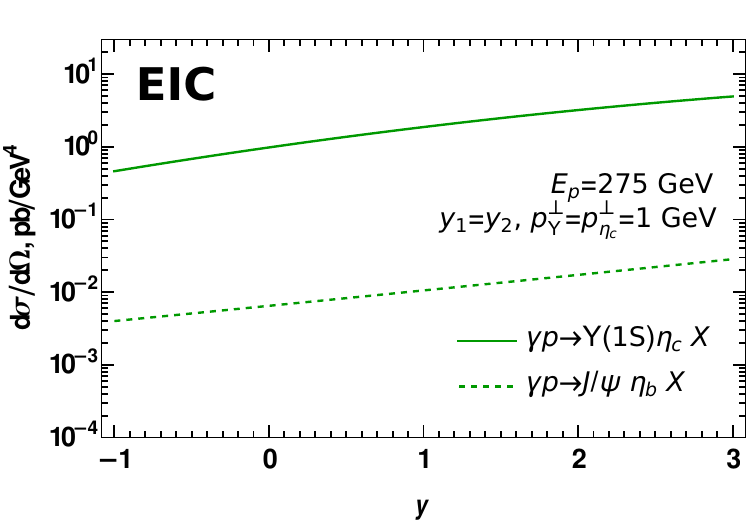}\includegraphics[width=9cm]{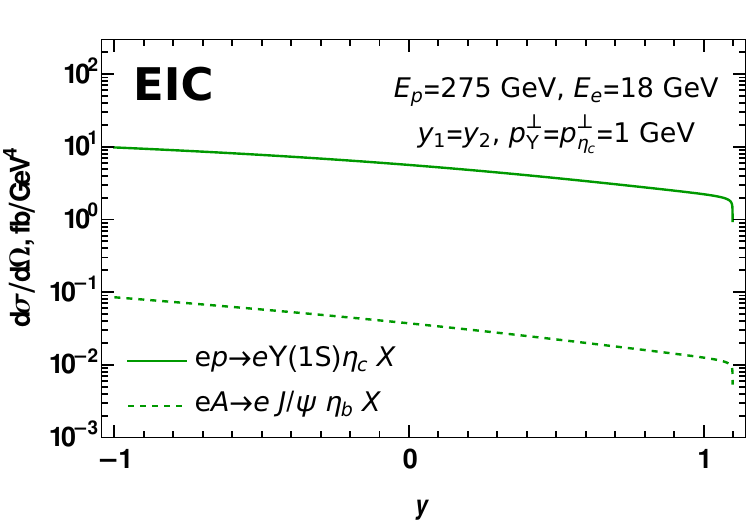}

\includegraphics[width=9cm]{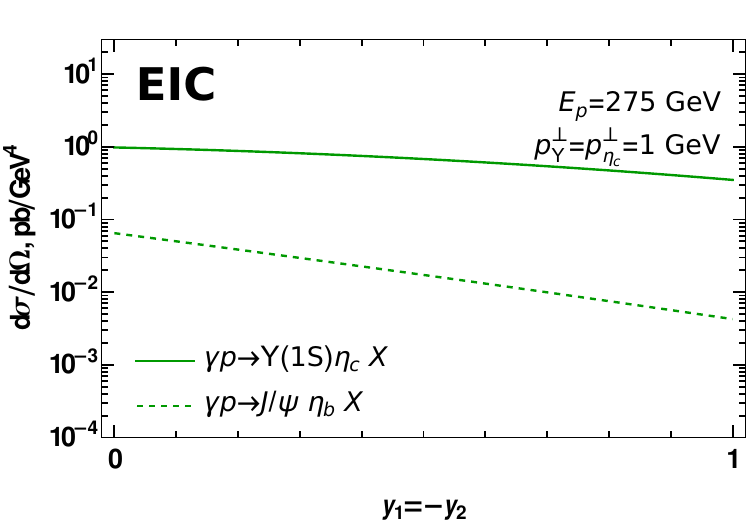}\includegraphics[width=9cm]{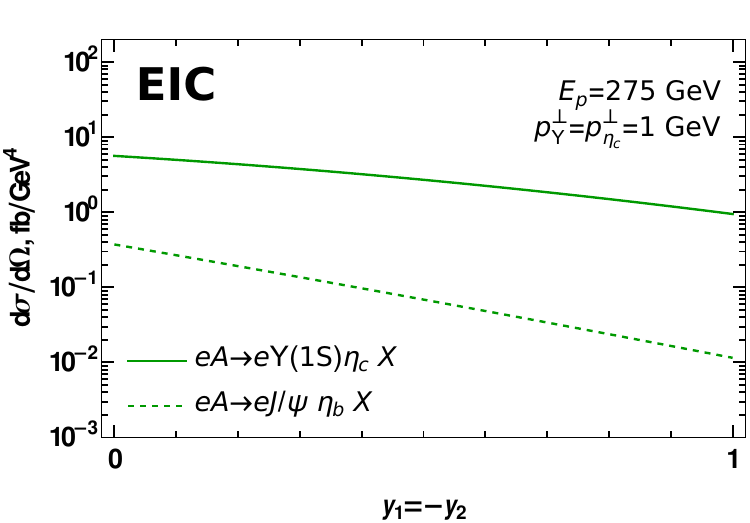}

\caption{\label{fig:Rapidities-EIC}The rapidity dependence of the cross-section
in the kinematics of the future Electron-Ion Collider. The left column
corresponds to the cross-section of the \textit{photo}production subprocess,
and the right column shows predictions for the cross-section of the
full electroproduction process $eA\to eAM_{1}M_{2}X$. Due to heavy
mass of the final state, the CGC approach is not applicable for $y\lesssim-1$,
and for $y\gtrsim1$ the flux of the equivalent photons is suppressed
due to leptonic factor (energy of the emitted virtual photon approaches
from below the energy of the electron). }
\end{figure}

Finally, in the Figures~\ref{fig:Rapidities-UPC},~\ref{fig:Rapidities-EIC}
we show the dependence on quarkonia rapidities in LHC and EIC kinematics.
The growth of the cross-section as a function of average rapidity
$Y=(y_{1}+y_{2})/2$ merely repeats the rapidity dependence (growth)
of the dipole scattering amplitude, which contributes directly and
indirectly via parametrization of quadrupole amplitude. In the second
row of the Figures~\ref{fig:Rapidities-UPC},~\ref{fig:Rapidities-EIC}
we show the dependence of the cross-section on the rapidity difference
$\Delta y$ between the two heavy mesons. For simplicity we made plots
for the quarkonia having opposite rapidities in the lab frame, $y_{1}=-y_{2}=\Delta y/2$.
The configurations with large rapidity difference correspond to highly
asymmetric sharing of the photon momentum between the quarks, for
which the wave function $\psi_{\bar{Q}Q\bar{Q}Q}^{(\gamma)}$ is strongly
suppressed. For this reason, the quarkonia are predominantly produced
with close rapidities. Since the variable $\Delta y$ may be related
to the invariant mass of the heavy quarkonia pairs~(\ref{eq:M12}),
this behavior implies suppression of the cross-section for large invariant
masses of charmonia-bottomonia pairs.


\section{Conclusions}

\label{sec:Conclusions}In this manuscript we analyzed in detail the\textit{
}inclusive photoproduction of heavy charmonia-bottomonia pairs. We
focused on quarkonia pairs with opposite $C$-parities ($\Upsilon\,\eta_{c},\,J/\psi\,\eta_{b}$),
which do not require exchange of quantum numbers in the $t$-channel,
and thus have the largest cross-sections. We found that the cross-section
is sensitive to bilinear superposition of dipole and quadrupole contributions,
which contribute in the same combination $S_{4}\left(\boldsymbol{x}_{1},\,\boldsymbol{x}_{2},\,\boldsymbol{x}_{3},\,\boldsymbol{x}_{4}\right)-S_{2}\left(\boldsymbol{x}_{1},\,\boldsymbol{x}_{2}\right)\otimes S_{2}\left(\boldsymbol{x}_{3},\,\boldsymbol{x}_{4}\right)$
as in inclusive \textit{single} quarkonia production~\cite{Kang:2013hta}.
Due to possibility to vary indepedently the momenta of both quarkonia,
the suggested process can allow to get better understanding of the
quadrupole scattering amplitude $S_{4}\left(\boldsymbol{x}_{1},\,\boldsymbol{x}_{2},\,\boldsymbol{x}_{3},\,\boldsymbol{x}_{4}\right)$.
We analyzed the role of the quadrupole scattering using parametrizations
available from the literature~\cite{Iancu:2011nj,Iancu:2011ns}
and found that numerically it gives the dominant contribution: its
omission decreases the cross-section by a factor two-ten depending
on the kinematics. The quadrupole term gives a pronounced dependence
on the angle between transverse momenta of the quarkonia, which is
almost negligible in its absence. 

Numerically, the cross-sections of the suggested processes are small,
but within the reach of the ultraperipheral collision experiments
at LHC and at the future Electron-Ion Collider. The smallness of the
cross-section happens due to heavy masses of charm and bottom quarks,
which control the sizes of the produced $\bar{c}c$ and $\bar{b}b$
pairs. The suggested mechanism also contributes to charmonia-charmonia
production, where its contribution is up to two orders of magnitude
larger than for charmonia-bottomonia. However, as explained in Section~\ref{subsec:Formalism},
in that channel there are additional contributions which have a significantly
more complicated structure and require a separate study.

\section*{Acknowledgments}

We thank our colleagues at UTFSM university for encouraging discussions.
% We also acknowledge a contribution of Prof. Ivan Schmidt, who suggested
% this topic of research, yet unfortunately passed away before the main
% results of this manuscript were obtained.
This research was partially
supported by Proyecto ANID PIA/APOYO AFB230003 (Chile) and Fondecyt
(Chile) grant 1220242. \textquotedbl Powered@NLHPC: This research
was partially supported by the supercomputing infrastructure of the
NLHPC (ECM-02)\textquotedbl .

\appendix

\section{Photon wave function $\psi_{\bar{Q}Q\bar{Q}Q}^{(\gamma)}$}

\label{sec:WF}The evaluation of the photon wave function follows
the standard light--cone rules formulated in~\cite{Lepage:1980fj,Brodsky:1997de}.
The result for the $\bar{Q}Q$ component is well-known in the literature~\cite{Bjorken:1970ah,Dosch:1996ss}
and is given by 
\begin{equation}
\Psi_{h\bar{h}}^{\lambda}\left(z,\,\boldsymbol{r}_{12},\,m_{q},a\right)=-\frac{2}{(2\pi)}\left[\left(z\delta_{\lambda,h}-(1-z)\delta_{\lambda,-h}\right)\delta_{h,-\bar{h}}i\boldsymbol{\varepsilon}_{\lambda}\cdot\nabla-\frac{m_{q}}{\sqrt{2}}\,{\rm sign}(h)\delta_{\lambda,h}\delta_{h,\bar{h}}\right]K_{0}\left(a\,\boldsymbol{r}_{12}\right).\label{eq:PsiSplitting}
\end{equation}
where $\lambda,h,\bar{h}$ are the helicities of the incoming photon
and outgoing quark, $\varepsilon^{\mu}(q)$ is the polarization vector
of the photon, $z$ is the fraction of the photon momentum carried
by the quark, and $\boldsymbol{r}_{12}$ is the transverse distance
between quark and antiquark. The wave function of the $\bar{Q}Q\bar{Q}Q$-component
can be expressed in terms of the wave function of $\bar{Q}Q$-component.
In what follows we will focus on the photoproduction kinematics, assuming
the transverse polarization of the incoming photons. We will use the
reference frame in which the photon momentum has only plus-component,
\begin{equation}
q\approx\left(q^{+},\,0,\,\boldsymbol{0}_{\perp}\right).
\end{equation}
so the polarization vector of the incoming photon is given by 
\begin{align}
\varepsilon_{T}^{\mu}(q) & \equiv\left(0,\,\frac{\boldsymbol{q}_{\perp}\cdot\boldsymbol{\boldsymbol{\varepsilon}}_{\gamma}}{q^{+}},\,\boldsymbol{\boldsymbol{\varepsilon}}_{\gamma}\right)\approx\left(0,\,0,\,\boldsymbol{\boldsymbol{\varepsilon}}_{\gamma}\right),\quad\boldsymbol{\boldsymbol{\varepsilon}}_{\gamma}=\frac{1}{\sqrt{2}}\left(\begin{array}{c}
1\\
\pm i
\end{array}\right),\quad\gamma=\pm1.\label{eq:eDef}
\end{align}

\begin{figure}
\includegraphics[scale=0.65]{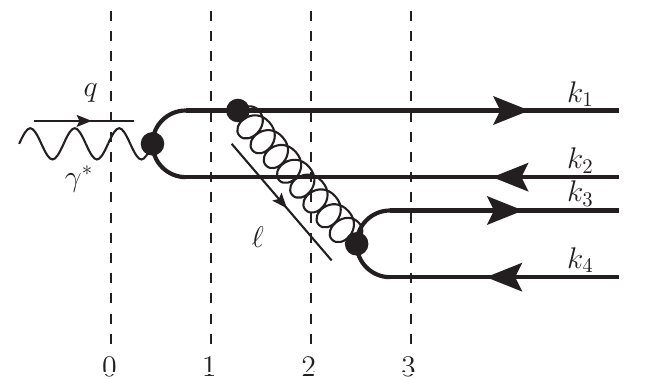}\includegraphics[scale=0.65]{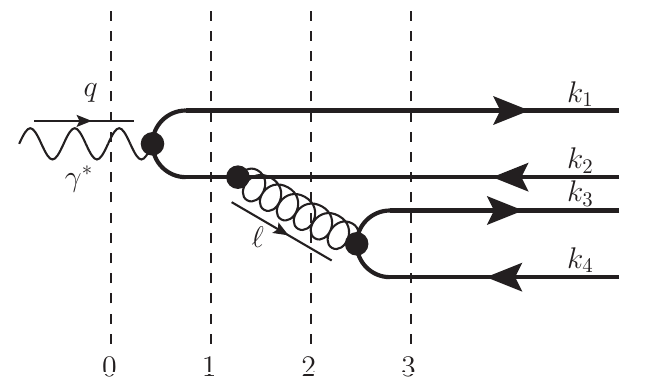}

\caption{\label{fig:Photoproduction-QQQQ}The leading order contribution to
the wave function $\psi_{\bar{Q}Q\bar{Q}Q}^{(\gamma)}$ defined in
the text. The momenta $k_{i}$ shown in the right-hand side are Fourier
conjugates of the coordinates $x_{i}$. It is implied that both diagrams
should be supplemented by all possible permutations of final state
quarks (see the text for more details).}
\end{figure}

In leading order over $\alpha_{s}$ the wave function obtains contributions
from the two diagrams shown in the Figure~(\ref{fig:Photoproduction-QQQQ}).
In what follows we will assume that the produced quark-antiquark pairs
have different flavors, and will use the notations $m_{1}$ for the
current mass of the quark line connected to a photon, and $m_{2}$
for the current masses of the quark-antiquark pair produced from the
virtual gluon. The evaluation of the diagrams follows the standard
rules of the light-cone perturbation theory~\cite{Lepage:1980fj,Brodsky:1997de}.
The detailed evaluation of the $\bar{Q}Q\bar{Q}Q$ component of the
photon wave function $\psi_{\bar{Q}Q\bar{Q}Q}^{(\gamma)}$ may be
found in~\cite{Andrade:2022rbn}, and here for the sake of completeness
we provide only the final result. This wave function can be represented as a sum
\begin{equation}
\psi_{\bar{Q}Q\bar{Q}Q}^{(\gamma)}=\psi_{\bar{Q}Q\bar{Q}Q}^{(\gamma,{\rm noninst})}+\psi_{\bar{Q}Q\bar{Q}Q}^{(\gamma,{\rm inst})}
\end{equation}
where the first and the second terms correspond to contributions of
non-instantaneous and instantaneous parts of propagators of all virtual
particles, and for the sake of brevity we omitted color and helicity
indices of heavy quarks ($c_{i}$ and $a_{i}$ respectively). The
non-instantaneous contribution is given by the sum 
\begin{equation}
\psi_{\bar{Q}Q\bar{Q}Q}^{(\gamma,{\rm noninst})}\left(\left\{ \alpha_{i},\,\boldsymbol{x}_{i}\right\} \right)=A\left(\left\{ \alpha_{i},\,\boldsymbol{x}_{i}\right\} \right)+B\left(\left\{ \alpha_{i},\,\boldsymbol{x}_{i}\right\} \right).\label{eq:PsiFull}
\end{equation}

where

\begin{align}
A\left(\left\{ \alpha_{i},\,\boldsymbol{r}_{i}\right\} \right) & =-\frac{2e_{q}\alpha_{s}\left(\mu\right)\,\left(t_{a}\right)_{c_{1}c_{2}}\otimes\left(t_{a}\right)_{c_{3}c_{4}}}{\pi^{3}\left(1-\alpha_{1}-\alpha_{2}\right)^{2}\sqrt{\alpha_{1}\alpha_{2}}\,}\int\frac{q_{1}dq_{1}\,k_{2}dk_{2}}{\frac{\bar{\alpha}_{2}q_{1}^{2}}{\alpha_{1}\left(1-\alpha_{1}-\alpha_{2}\right)}+\frac{m_{1}^{2}\left(\alpha_{1}+\alpha_{2}\right)}{\alpha_{1}\alpha_{2}}+\frac{k_{2}^{2}}{\alpha_{2}\bar{\alpha}_{2}}}\times\label{eq:Amp_p_explicit-4-4}\\
 & \times\frac{1}{k_{2}^{2}+m_{1}^{2}}\sqrt{\frac{\alpha_{2}}{\alpha_{1}}}\left[\left(\alpha_{2}\delta_{\gamma,a_{2}}-\bar{\alpha}_{2}\delta_{\gamma,-a_{2}}\right)\left(\bar{\alpha}_{2}\delta_{\lambda,\,a_{1}}+\alpha_{1}\delta_{\lambda,-a_{1}}\right)\delta_{a_{1},-a_{2}}\times\right.\nonumber \\
 & \times\left(\boldsymbol{n}_{2,134}\cdot\boldsymbol{\varepsilon}_{\gamma}\right)\left(\boldsymbol{n}_{1,34}\cdot\boldsymbol{\varepsilon}_{\lambda}^{*}\right)k_{2}\,J_{1}\left(k_{2}\left|\boldsymbol{x}_{2}-\boldsymbol{b}_{134}\right|\right)q_{1}J_{1}\left(q_{1}\left|\boldsymbol{x}_{1}-\boldsymbol{b}_{34}\right|\right)+\nonumber \\
 & +\frac{m_{q}^{2}}{2}\,\delta_{\lambda,-a_{1}}\delta_{\gamma,a_{2}}\delta_{a_{1},-a_{2}}J_{0}\left(k_{2}\left|\boldsymbol{x}_{2}-\boldsymbol{b}_{134}\right|\right)J_{0}\left(q_{1}\left|\boldsymbol{x}_{1}-\boldsymbol{b}_{34}\right|\right)\frac{\left(1-\alpha_{1}-\alpha_{2}\right)^{2}}{1-\alpha_{2}}\nonumber \\
 & -\frac{im_{q}}{\sqrt{2}}\,{\rm sign}\left(a_{2}\right)\delta_{\gamma,a_{2}}\delta_{a_{1},a_{2}}\left(\bar{\alpha}_{2}\delta_{\lambda,\,a_{1}}+\alpha_{1}\delta_{\lambda,-a_{1}}\right)\times\nonumber \\
 & \times\boldsymbol{n}_{1,34}\cdot\boldsymbol{\varepsilon}_{\lambda}^{*}q_{1}J_{1}\left(q_{1}\left|\boldsymbol{x}_{1}-\boldsymbol{b}_{34}\right|\right)J_{0}\left(k_{2}\left|\boldsymbol{x}_{2}-\boldsymbol{b}_{134}\right|\right)\nonumber \\
 & -\frac{im_{q}}{\sqrt{2}}{\rm sign}\left(a_{1}\right)\delta_{\lambda,-a_{1}}\left(\alpha_{2}\delta_{\gamma,a_{2}}-\bar{\alpha}_{2}\delta_{\gamma,-a_{2}}\right)\delta_{a_{1},a_{2}}\frac{\left(1-\alpha_{1}-\alpha_{2}\right)^{2}}{1-\alpha_{2}}\times\nonumber \\
 & \times\left.\left(\boldsymbol{n}_{2,134}\cdot\boldsymbol{\varepsilon}_{\gamma}\right)k_{2}\,J_{1}\left(k_{2}\left|\boldsymbol{x}_{2}-\boldsymbol{b}_{134}\right|\right)J_{0}\left(q_{1}\left|\boldsymbol{x}_{1}-\boldsymbol{b}_{34}\right|\right)\right]\times\nonumber \\
 & \times\Psi_{a_{3},a_{4}}^{-\lambda}\left(\frac{\alpha_{3}}{\alpha_{3}+\alpha_{4}},\,\boldsymbol{r}_{34},\,m_{2},\,\sqrt{m_{2}^{2}+\frac{\alpha_{3}\alpha_{4}}{\alpha_{3}+\alpha_{4}}\left[\frac{\bar{\alpha}_{2}q_{1}^{2}}{\alpha_{1}\left(1-\alpha_{1}-\alpha_{2}\right)}+\frac{m_{1}^{2}\left(\alpha_{1}+\alpha_{2}\right)}{\alpha_{1}\alpha_{2}}+\frac{k_{2}^{2}}{\alpha_{2}\bar{\alpha}_{2}}\right]}\right),\nonumber
\end{align}
\begin{equation}
\Psi_{h\bar{h}}^{\lambda}\left(z,\,\boldsymbol{r}_{12},\,m_{q},a\right)=-\frac{2}{(2\pi)}\left[\left(z\delta_{\lambda,h}-(1-z)\delta_{\lambda,-h}\right)\delta_{h,-\bar{h}}i\boldsymbol{\varepsilon}_{\lambda}\cdot\nabla-\frac{m_{q}}{\sqrt{2}}\,{\rm sign}(h)\delta_{\lambda,h}\delta_{h,\bar{h}}\right]K_{0}\left(a\,\boldsymbol{r}\right).\label{eq:PsiSplitting-1}
\end{equation}
and 
\[
B\left(\alpha_{1},\,\boldsymbol{x}_{1},\,\alpha_{2},\,\boldsymbol{x}_{2},\,\alpha_{3},\,\boldsymbol{x}_{3},\,\alpha_{4},\,\boldsymbol{x}_{4}\right)=-A\left(\alpha_{2},\,\boldsymbol{x}_{2},\,\alpha_{1},\,\boldsymbol{x}_{1},\,\alpha_{4},\,\boldsymbol{x}_{4},\,\alpha_{3},\,\boldsymbol{x}_{3}\right).
\]
We also introduced a shorthand notation 
\begin{equation}
\boldsymbol{b}_{j_{1}...j_{n}}=\frac{\alpha_{j_{1}}\boldsymbol{x}_{j_{1}}+...+\alpha_{j_{n}}\boldsymbol{x}_{j_{n}}}{\alpha_{j_{1}}+...+\alpha_{j_{n}}}
\end{equation}
for the center of mass position of the $n$ partons $j_{1},\,...j_{n}$,
and 
\begin{equation}
\boldsymbol{n}_{i,j_{1}...j_{n}}=\frac{\boldsymbol{x}_{i}-\boldsymbol{b}_{j_{1}...j_{n}}}{\left|\boldsymbol{x}_{i}-\boldsymbol{b}_{j_{1}...j_{n}}\right|}
\end{equation}
for a unit vector pointing from quark $i$ towards the center-of-mass
of the system of quarks $j_{1}...j_{n}$. The variables $\boldsymbol{r}_{1}-\boldsymbol{b}_{34}$
and $\boldsymbol{r}_{2}-\boldsymbol{b}_{34}$ physically have the
meaning of the relative distance between the recoil quark or antiquark
and the emitted gluon. Similarly, the variables $\boldsymbol{r}_{1}-\boldsymbol{b}_{234}$
and $\boldsymbol{r}_{2}-\boldsymbol{b}_{134}$ might be interpreted
as the size of $\bar{Q}Q$ pair produced right after splitting of
the incident photon. For the instantaneous contributions it is possible
to get 
\begin{align}
\psi_{\bar{Q}Q\bar{Q}Q}^{(\gamma,{\rm inst.})}\left(\left\{ \alpha_{i},\,\boldsymbol{r}_{i}\right\} \right) & =A_{g}\left(\left\{ \alpha_{i},\,\boldsymbol{r}_{i}\right\} \right)+B_{g}\left(\left\{ \alpha_{i},\,\boldsymbol{r}_{i}\right\} \right)+\label{eq:PsiFull-1}\\
 & +A_{q}\left(\left\{ \alpha_{i},\,\boldsymbol{r}_{i}\right\} \right)+B_{q}\left(\left\{ \alpha_{i},\,\boldsymbol{r}_{i}\right\} \right),\nonumber 
\end{align}
where the subscript indices $q,g$ in the right-hand side denote the
parton propagator which should be taken instantaneous ($q$ for quark,
$g$ for gluon), and

\begin{align}
A_{g}\left(\left\{ \alpha_{i},\,\boldsymbol{r}_{i}\right\} \right) & =-\frac{e_{q}\alpha_{s}(m_{Q})\,\left(t_{a}\right)_{c_{1}c_{2}}\otimes\left(t_{a}\right)_{c_{3}c_{4}}}{\pi^{4}\left(1-\alpha_{1}-\alpha_{2}\right)^{3}\,}\int q_{1}dq_{1}\,k_{2}dk_{2}J_{0}\left(q_{1}\left|\boldsymbol{r}_{1}-\boldsymbol{b}_{34}\right|\right)\times\\
 & \times\frac{1}{\boldsymbol{k}_{2\perp}^{2}+m_{1}^{2}}\left[\left(\alpha_{2}\delta_{\gamma,a_{1}}-\bar{\alpha}_{2}\delta_{a_{1},-\gamma}\right)\delta_{a_{1},-a_{2}}i\boldsymbol{n}_{2,134}\cdot\boldsymbol{\varepsilon}_{\gamma}k_{2}J_{1}\left(k_{2}\left|\boldsymbol{r}_{2}-\boldsymbol{b}_{134}\right|\right)\right.\nonumber \\
 & \left.+\frac{m_{q}}{\sqrt{2}}\,{\rm sign}\left(a_{1}\right)\delta_{\gamma,a_{1}}\delta_{a_{1},a_{2}}J_{0}\left(k_{2}\left|\boldsymbol{r}_{2}-\boldsymbol{b}_{134}\right|\right)\right]\alpha_{3}\alpha_{4}\delta_{a_{3},-a_{4}}K_{0}\left(a_{34}\,\boldsymbol{r}_{34}\right).\nonumber 
\end{align}
\begin{align}
A_{q}\left(\left\{ \alpha_{i},\,\boldsymbol{r}_{i}\right\} \right) & =-\frac{e_{q}\alpha_{s}\left(m_{q}\right)\,\left(t_{a}\right)_{c_{1}c_{2}}\otimes\left(t_{a}\right)_{c_{3}c_{4}}}{2\pi^{4}\left(1-\alpha_{1}-\alpha_{2}\right)^{2}\bar{\alpha}_{2}}\delta_{a_{1},-a_{2}}\delta_{\gamma,-a_{1}}\int q_{1}dq_{1}\,k_{2}dk_{2}\frac{J_{0}\left(q_{1}\left|\boldsymbol{r}_{1}-\boldsymbol{b}_{34}\right|\right)J_{0}\left(k_{2}\left|\boldsymbol{r}_{2}-\boldsymbol{b}_{134}\right|\right)}{D_{2}\left(\alpha_{1},\boldsymbol{k}_{1};\alpha_{2},\,\boldsymbol{k}_{2}\right)}\times\\
 & \times\left[-\left(\alpha_{3}\delta_{-\gamma,a_{3}}-\alpha_{4}\delta_{\gamma,a_{3}}\right)\delta_{a_{3},-a_{4}}i\boldsymbol{\varepsilon}_{\gamma}\cdot\boldsymbol{n}_{34}a_{34}K_{1}\left(a_{34}\,\boldsymbol{r}_{34}\right)-\frac{m_{q}(\alpha_{3}+\alpha_{4})}{\sqrt{2}}\,{\rm sign}(a_{3})\delta_{\gamma,-a_{3}}\delta_{a_{3},a_{4}}K_{0}\left(a_{34}\,\boldsymbol{r}_{34}\right)\right]\nonumber \\
 & a_{34}\left(q_{1},\,k_{2}\right)\equiv\sqrt{m_{2}^{2}+\frac{\alpha_{3}\alpha_{4}}{\alpha_{3}+\alpha_{4}}\left[\frac{\bar{\alpha}_{2}q_{1}^{2}}{\alpha_{1}\left(1-\alpha_{1}-\alpha_{2}\right)}+\frac{m_{1}^{2}\left(\alpha_{1}+\alpha_{2}\right)}{\alpha_{1}\alpha_{2}}+\frac{k_{2}^{2}}{\alpha_{2}\bar{\alpha}_{2}}\right]}
\end{align}
and the functions $B_{q},\,B_{g}$ can be obtained from $A_{q},A_{g}$
using 
\begin{equation}
B_{i}\left(\alpha_{1},\,\boldsymbol{x}_{1},\,\alpha_{2},\,\boldsymbol{x}_{2},\,\alpha_{3},\,\boldsymbol{x}_{3},\,\alpha_{4},\,\boldsymbol{x}_{4}\right)=-A_{i}\left(\alpha_{2},\,\boldsymbol{x}_{2},\,\alpha_{1},\,\boldsymbol{x}_{1},\,\alpha_{4},\,\boldsymbol{x}_{4},\,\alpha_{3},\,\boldsymbol{x}_{3}\right),\quad i=q,g.
\end{equation}
We can observe that in all terms the dependence on color structure
is encoded in a common factor $\sim\left(t_{a}\right)_{c_{1}c_{2}}\otimes\left(t_{a}\right)_{c_{3}c_{4}}$,
which drastically simplifies evaluations and allows to cast the result
in the form of a convolution of wave functions and dipole amplitude.

\section{Evaluation of the multipole correlator}

\label{sec:Multipole}In this section we evaluate the multipole correlator
\begin{equation}
\mathcal{A}_{c_{1}'c_{2}'c_{3}'c_{4}'}^{c_{1}c_{2}c_{3}c_{4}}\left(\boldsymbol{x}_{1},\boldsymbol{y}_{1},...,\boldsymbol{x}_{4},\boldsymbol{y}_{4}\right)=\left\langle \prod_{k=1}^{4}\left[U^{\dagger}\left(\boldsymbol{x}_{k}\right)U\left(\boldsymbol{y}_{k}\right)\right]_{c_{a}'\,}^{c_{a}}\right\rangle \label{eq:Adef}
\end{equation}
where $c_{a},c_{a}'$ are the color indices in fundamental representation
of the color group, and angular brackets stand for the color averaging
$\int\mathcal{D}\mu\left[U_{\boldsymbol{x}}\right]$. As explained
in the text, this amplitude appears when we consider production of
two color singlet quarkonia; for color octet channel the matrices
$U^{\dagger},\,U$ are separated by additional color generators $t^{a}$.
Previously, similar correlator has been evaluated exactly in~\cite{Kovner:2017ssr,Kovner:2018vec}
in the context of studies of inclusive 3-quark production; for the
4-quark production the result was provided in implicit form (convoluted
with hard amplitudes) in the large-$N_{c}$ limit, when all multipoles
are expressed via dipole amplitudes. For the sake of completeness,
below we provide a complete result for the amplitude.

Since the target is color singlet, it is expected that color averaging
of the right-hand side should eventually include a sum over all possible
color singlet irreducible representations which appear in direct product
of $U^{\dagger}U$ matrices, namely
\begin{equation}
\mathcal{A}_{c_{1}'c_{2}'c_{3}'c_{4}'}^{c_{1}c_{2}c_{3}c_{4}}\left(\boldsymbol{x}_{1},\boldsymbol{y}_{1},...,\boldsymbol{x}_{4},\boldsymbol{y}_{4}\right)=\sum_{\boldsymbol{p}}\delta_{c_{p_{1}}'}^{c_{1}}\delta_{c_{p_{2}}'}^{c_{2}}\delta_{c_{p_{3}}'}^{c_{3}}\delta_{c_{p_{4}}'}^{c_{4}}a_{\boldsymbol{p}}\left(\boldsymbol{x}_{1},\boldsymbol{y}_{p_{1}},...,\boldsymbol{x}_{4},\boldsymbol{y}_{p_{4}}\right)\label{eq:Gen24}
\end{equation}
where summation is done over all possible permutations $\boldsymbol{p}$
of numbers (1,2,3,4), and we use a shorthand notation $p_{k}$ for
the $k$-th element of permutation $\boldsymbol{p}$. Overall, expression~(\ref{eq:Gen24})
includes $4!=24$ unknown functions $a_{\boldsymbol{p}}(...)$. In
order to fix them, we may combine~(\ref{eq:Adef},\ref{eq:Gen24})
and multiply both parts by
\[
\delta_{c_{1}}^{c_{\ell_{1}}'}\delta_{c_{2}}^{c_{\ell_{2}}'}\delta_{c_{3}}^{c_{\ell_{3}}'}\delta_{c_{4}}^{c_{\ell_{4}}'}
\]
where $\boldsymbol{\ell}=\left(\ell_{1},\ell_{2},\ell_{3},\ell_{4}\right)$
is some arbitrary fixed permutaton. This yields a system of linear
inhomogeneous equations
\begin{align}
 & \mathcal{M}_{\boldsymbol{\ell},\,\boldsymbol{p}}\cdot a_{\boldsymbol{p}}=B_{\boldsymbol{\ell}}\label{eq:MaB}
\end{align}
where 
\begin{align}
 & \mathcal{M}_{\boldsymbol{\ell},\,\boldsymbol{p}}=\delta_{c_{p_{1}}}^{c_{\ell_{1}}}\delta_{c_{p_{2}}}^{c_{\ell_{2}}}\delta_{c_{p_{3}}}^{c_{\ell_{3}}}\delta_{c_{p_{4}}}^{c_{\ell_{4}}},\\
 & B_{\boldsymbol{\ell}}=\left\langle \prod_{k=1}^{4}\left[U^{\dagger}\left(\boldsymbol{x}_{k}\right)U\left(\boldsymbol{y}_{k}\right)\right]_{c_{a}'\,}^{c_{a}}\right\rangle \delta_{c_{1}}^{c_{\ell_{1}}'}\delta_{c_{2}}^{c_{\ell_{2}}'}\delta_{c_{3}}^{c_{\ell_{3}}'}\delta_{c_{4}}^{c_{\ell_{4}}'}\label{eq:B}
\end{align}
All the elements of the matrix $\mathcal{M}_{\boldsymbol{\ell},\,\boldsymbol{p}}$have
very simple structure and are given just by $N_{c}^{k}$, where $1\le k\le4$
is the number of disjoint cycles in permutation $\boldsymbol{\ell}\cdot\boldsymbol{p}$.
For example, if $\boldsymbol{\ell}_{0}=(1,2,3,4)$ and $\boldsymbol{p}_{0}=(2,1,4,3)$,
then the corresponding element 
\begin{equation}
\mathcal{M}_{\boldsymbol{\ell}_{0},\,\boldsymbol{p}_{0}}=\delta_{c_{2}}^{c_{1}}\delta_{c_{1}}^{c_{2}}\,\delta_{c_{4}}^{c_{3}}\delta_{c_{3}}^{c_{4}}=N_{c}^{2}.
\end{equation}
A contraction of $\delta$-symbols in the right-hand side of~(\ref{eq:B})
allows to express all elements $B_{\boldsymbol{\ell}}$ in terms of
color singlet multitrace operators. For example, for permutations
$\boldsymbol{p}_{0}=(2,1,4,3)$, $\boldsymbol{p}_{1}=(2,3,1,4)$ and
$\boldsymbol{p}_{2}=(2,3,4,1)$ we may get respectively
\begin{align}
B_{\boldsymbol{p}_{0}} & =\left\langle {\rm tr}\left[U^{\dagger}\left(\boldsymbol{x}_{1}\right)U\left(\boldsymbol{y}_{1}\right)U^{\dagger}\left(\boldsymbol{x}_{2}\right)U\left(\boldsymbol{y}_{2}\right)\right]{\rm tr}\left[U^{\dagger}\left(\boldsymbol{x}_{3}\right)U\left(\boldsymbol{y}_{3}\right)U^{\dagger}\left(\boldsymbol{x}_{4}\right)U\left(\boldsymbol{y}_{4}\right)\right]\right\rangle \\
B_{\boldsymbol{p}_{1}} & =\left\langle {\rm tr}\left[U^{\dagger}\left(\boldsymbol{x}_{1}\right)U\left(\boldsymbol{y}_{1}\right)U^{\dagger}\left(\boldsymbol{x}_{2}\right)U\left(\boldsymbol{y}_{2}\right)U^{\dagger}\left(\boldsymbol{x}_{3}\right)U\left(\boldsymbol{y}_{3}\right)\right]{\rm tr}\left[U^{\dagger}\left(\boldsymbol{x}_{4}\right)U\left(\boldsymbol{y}_{4}\right)\right]\right\rangle ,\\
B_{\boldsymbol{p}_{2}} & =\left\langle {\rm tr}\left[U^{\dagger}\left(\boldsymbol{x}_{1}\right)U\left(\boldsymbol{y}_{1}\right)U^{\dagger}\left(\boldsymbol{x}_{2}\right)U\left(\boldsymbol{y}_{2}\right)U^{\dagger}\left(\boldsymbol{x}_{3}\right)U\left(\boldsymbol{y}_{3}\right)U^{\dagger}\left(\boldsymbol{x}_{4}\right)U\left(\boldsymbol{y}_{4}\right)\right]\right\rangle ,
\end{align}
which includes in the right-hand side quadrupole, sextupole and octupole
contributions. The solution of nonlinear system~(\ref{eq:MaB}) is
straightforward and was done with \textit{Mathematica} (and FeynCalc
package~\cite{FeynCalc1,FeynCalc2} for color group algebra). The
result for the constants $a_{\boldsymbol{p}}$ is very lengthy if
written in terms of conventional notations $S_{2n}\left(x_{1}\right)$,
for this reason, for the sake of brevity we will introduce shorthand
notations 
\begin{align}
[i] & =S_{2}\left(\boldsymbol{x}_{i},\boldsymbol{y}_{i}\right),\quad[i,j]=S_{4}\left(\boldsymbol{x}_{i},\boldsymbol{y}_{i},\boldsymbol{x}_{j},\boldsymbol{y}_{j}\right),\quad[i,j,k]=S_{6}\left(\boldsymbol{x}_{i},\boldsymbol{y}_{i},\boldsymbol{x}_{j},\boldsymbol{y}_{j},\boldsymbol{x}_{k},\boldsymbol{y}_{k}\right),\\
 & [i,j,k,\ell]=S_{8}\left(\boldsymbol{x}_{i},\boldsymbol{y}_{i},\boldsymbol{x}_{j},\boldsymbol{y}_{j},\boldsymbol{x}_{k},\boldsymbol{y}_{k},\boldsymbol{x}_{\ell},\boldsymbol{y}_{\ell}\right),
\end{align}
where each of the indices $i,j,k,\ell$ may take integer values between
1 and 4. In these notations the amplitude $\mathcal{A}$ is written
as
\begin{align}
\mathcal{A} & _{c_{1}'c_{2}'c_{3}'c_{4}'}^{c_{1}c_{2}c_{3}c_{4}}\left(\boldsymbol{x}_{1},\boldsymbol{y}_{1},...,\boldsymbol{x}_{4},\boldsymbol{y}_{4}\right)=\frac{\delta_{c_{1}'}^{c_{1}}\delta_{c_{2}'}^{c_{2}}\delta_{c_{3}'}^{c_{3}}\delta_{c_{4}'}^{c_{4}}}{N_{c}^{2}\left(N_{c}^{2}-7\right)^{2}-36}\left[\left(4-N_{c}^{2}\right)N_{c}^{2}\left(\frac{}{}\left\langle [1][2][3,4]+[1][3][2,4]\right\rangle \right.\right.\label{eq:AmpFull}\\
 & +\left.\left\langle [1][4][2,3]+[2][4][1,3]+[2][3][1,4]+[3][4][1,2]\right\rangle \frac{}{}\right)+\left(N_{c}^{2}+6\right)\left\langle [1,2][3,4]+[1,4][2,3]+[1,3][2,4]\right\rangle \nonumber \\
 & +\left(2N_{c}^{2}-3\right)\left\langle [1][2,3,4]+[1][2,4,3]+[2][1,3,4]+[2][1,4,3]+[3][1,4,2]+[3][1,2,4]+[4][1,2,3]+[4][1,3,2]\right\rangle \nonumber \\
 & -5\left\langle [1,2,3,4]+[1,2,4,3]+[1,3,2,4]+[1,3,4,2]+[1,4,2,3]+[1,4,3,2]\right\rangle +\left.\left(N_{c}^{4}-8N_{c}^{2}+6\right)N_{c}^{2}\left\langle [1][2][3][4]\right\rangle \frac{}{}\right]\nonumber \\
+ & \frac{\delta_{c_{1}'}^{c_{1}}\delta_{c_{2}'}^{c_{2}}\delta_{c_{4}'}^{c_{3}}\delta_{c_{3}'}^{c_{4}}}{N_{c}^{2}\left(N_{c}^{2}-7\right)^{2}-36}\left[\frac{}{}\left(4-N_{c}^{2}\right)N_{c}^{3}[1][2][3][4]+N_{c}(2N_{c}^{2}-3)\left\langle [1][3][2,4]+[1][4][2,3]+[2][3][1,4]+[2][4][1,3]\right\rangle +\right.\nonumber \\
 & +\left(4-N_{c}^{2}\right)N_{c}\left\langle [1][2,3,4]+[1][2,4,3]+[2][1,3,4]+[2][1,4,3]+[1,2][3,4]\right\rangle +\left(N_{c}^{4}-8N_{c}^{2}+6\right)N_{c}\left\langle [1][2][3,4]\right\rangle \nonumber \\
 & +N_{c}\left(N_{c}^{2}+6\right)\langle[3][4][1,2]\rangle-5N_{c}\left\langle [1,4][2,3]+[1,3][2,4]+[3][1,2,4]+[3][1,4,2]+[4][1,2,3]+[4][1,3,2]\right\rangle \nonumber \\
 & +\left.\frac{}{}\left(2N_{c}-\frac{3}{N_{c}}\right)\left\langle [1,2,3,4]+[1,2,4,3]+[1,3,4,2]+[1,4,3,2]\right\rangle +\left(N_{c}+\frac{6}{N_{c}}\right)\left\langle [1,3,2,4]+[1,4,2,3]\right\rangle \right]\nonumber \\
+ & \frac{\delta_{c_{1}'}^{c_{1}}\delta_{c_{3}'}^{c_{2}}\delta_{c_{2}'}^{c_{3}}\delta_{c_{4}'}^{c_{4}}}{N_{c}\left(N_{c}^{2}\left(N_{c}^{2}-7\right)^{2}-36\right)}\left[\frac{}{}\left(4-N_{c}^{2}\right)N_{c}^{4}\left\langle [1][2][3][4]\right\rangle +\left(N_{c}^{4}-8N_{c}^{2}+6\right)N_{c}^{2}\left\langle [1][2,3][4]\right\rangle \right.+\nonumber \\
 & +\left(N_{c}^{2}+6\right)N_{c}^{2}\left\langle [1,4][2][3]\right\rangle +\left(4-N_{c}^{2}\right)N_{c}^{2}\left\langle [1][2,4,3]+[1][2,3,4]+[1,2,3][4]+[1,3,2][4]+[1,4][2,3]\right\rangle \nonumber \\
 & +\left(2N_{c}^{2}-3\right)N_{c}^{2}\left\langle [1][2][3,4]+[1,2][3][4]+[2,4][1][3]+[1,3][2][4]\right\rangle \nonumber \\
 & -5N_{c}^{2}\left\langle [1,2][3,4]+[1,2,4][3]+[1,3,4][2]+[1,4,2][3]+[1,4,3][2]+[1,3][2,4]\right\rangle \nonumber \\
 & \left.+\left(2N_{c}^{2}-3\right)\left\langle [1,2,3,4]+[1,3,2,4]+[1,4,2,3]+[1,4,3,2]\right\rangle +\left(N_{c}^{2}+6\right)\left\langle [1,2,4,3]+[1,3,4,2]\right\rangle \frac{}{}\right]\nonumber \\
+ & \frac{\delta_{c_{1}'}^{c_{1}}\delta_{c_{3}'}^{c_{2}}\delta_{c_{4}'}^{c_{3}}\delta_{c_{2}'}^{c_{4}}}{\left(N_{c}^{2}\left(N_{c}^{2}-7\right)^{2}-36\right)}\left[\frac{}{}\left(2N_{c}^{2}-3\right)N_{c}^{2}\left\langle [1][2][3][4]\right\rangle +\left(4-N_{c}^{2}\right)N_{c}^{2}\left\langle [1][2,4][3]+[1][2][3,4]+[1][2,3][4]\right\rangle \right.\nonumber \\
 & +\left\langle [1][2,3,4]\right\rangle \left(N_{c}^{4}-8N_{c}^{2}+6\right)\nonumber \\
 & +\left(2N_{c}^{2}-3\right)\left\langle [1,2][3,4]+[1,3][2,4]+[1,4][2,3]+[1,2,3][4]+[1][2,4,3]+[1,3,4][2]+[1,4,2][3]\right\rangle \nonumber \\
 & -5N_{c}^{2}\left\langle [1,2][3][4]+[1,3][2][4]+[2][3][1,4]\right\rangle +\left(4-N_{c}^{2}\right)\left\langle [1,2,3,4]+[1,3,4,2]+[1,4,2,3]\right\rangle \nonumber \\
 & +\left.\left(N_{c}^{2}+6\right)\left\langle [1,2,4][3]+[1,3,2][4]+[1,4,3][2]\right\rangle -5\left\langle [1,2,4,3]+[1,3,2,4]+[1,4,3,2]\right\rangle \frac{}{}\right]\nonumber \\
+ & \frac{\delta_{c_{1}'}^{c_{1}}\delta_{c_{4}'}^{c_{2}}\delta_{c_{2}'}^{c_{3}}\delta_{c_{3}'}^{c_{4}}}{N_{c}^{2}\left(N_{c}^{2}-7\right)^{2}-36}\left[\frac{}{}\left\langle [1][2][3][4]\right\rangle \left(2N_{c}^{2}-3\right)N_{c}^{2}+\left\langle [2,4,3][1]\right\rangle \left(N_{c}^{4}-8N_{c}^{2}+6\right)+\right.\nonumber \\
 & +\left(4-N_{c}^{2}\right)N_{c}^{2}\left\langle [1][2][3,4]+[1][2,3][4]+[1][2,4][3]\right\rangle +\left(4-N_{c}^{2}\right)\left\langle [1,2,4,3]+[1,3,2,4]+[1,4,3,2]\right\rangle \nonumber \\
 & +\left(2N_{c}^{2}-3\right)\left\langle [2,3,4][1]+[1,2,4][3]+[1,3][2,4]+[1,4][2,3]+[1,3,2][4]+[1,2][3,4]+[1,4,3][2]\right\rangle \nonumber \\
 & +\left(N_{c}^{2}+6\right)\left\langle [1,2,3][4]+[1,3,4][2]+[1,4,2][3]\right\rangle -5N_{c}^{2}\left\langle [1,2][3][4]+[1,3][2][4]+[1,4][2][3]\right\rangle \nonumber \\
 & \left.\frac{}{}-5\left\langle [1,2,3,4]+[1,3,4,2]+[1,4,2,3]\right\rangle \right]\nonumber \\
+ & \frac{\delta_{c_{1}'}^{c_{1}}\delta_{c_{4}'}^{c_{2}}\delta_{c_{3}'}^{c_{3}}\delta_{c_{2}'}^{c_{4}}}{N_{c}\left(N_{c}^{2}\left(N_{c}^{2}-7\right)^{2}-36\right)}\left[\left\langle [1][2][3][4]\right\rangle \left(4-N_{c}^{2}\right)N_{c}^{4}+\left(2N_{c}^{2}-3\right)N_{c}^{2}\left\langle [1,2][3][4]+[1][2,3][4]+[1][2][3,4]+[1,4][2][3]\right\rangle \frac{}{}\right.\nonumber \\
 & +\left\langle [2,4][1][3]\right\rangle \left(N_{c}^{4}-8N_{c}^{2}+6\right)N_{c}^{2}+\left(4-N_{c}^{2}\right)N_{c}^{2}\left\langle [2,3,4][1]+[2,4,3][1]+[1,3][2,4]+[1,2,4][3]+[1,4,2][3]\right\rangle \nonumber \\
 & -5N_{c}^{2}\left\langle [1,2][3,4]+[1,4][2,3]+[1,2,3][4]+[1,3,4][2]+[1,3,2][4]+[1,4,3][2]\right\rangle +\left(N_{c}^{2}+6\right)\left\langle [1,2,3,4]+[1,4,3,2]\right\rangle \nonumber \\
 & \left.+\left(N_{c}^{2}+6\right)N_{c}^{2}\left\langle [1,3][2][4]\right\rangle +\left(2N_{c}^{2}-3\right)\left\langle [1,2,4,3]+[1,3,2,4]+[1,3,4,2]+[1,4,2,3]\right\rangle \frac{}{}\right]\nonumber \\
+ & \frac{\delta_{c_{2}'}^{c_{1}}\delta_{c_{1}'}^{c_{2}}\delta_{c_{3}'}^{c_{3}}\delta_{c_{4}'}^{c_{4}}}{N_{c}\left(N_{c}^{2}\left(N_{c}^{2}-7\right)^{2}-36\right)}\left[\frac{}{}-\left(N_{c}^{2}-4\right)N_{c}^{4}\left\langle [1][2][3][4]\right\rangle +\left(N_{c}^{2}+6\right)N_{c}^{2}\left\langle [1][2][3,4]\right\rangle +\right.\nonumber \\
 & +\left(2N_{c}^{2}-3\right)N_{c}^{2}\left\langle [1][2,3][4]+[2,4][1][3]+[1,3][2][4]+[1,4][2][3]\right\rangle +\left(N_{c}^{2}+6\right)\left\langle [1,3,2,4]+[1,4,2,3]\right\rangle \nonumber \\
 & +N_{c}^{2}\left(4-N_{c}^{2}\right)\left\langle [1,2][3,4]+[1,2,3][4]+[1,2,4][3]+[1,3,2][4]+[1,4,2][3]\right\rangle \nonumber \\
 & +\left\langle [1,2][3][4]\right\rangle \left(N_{c}^{4}-8N_{c}^{2}+6\right)N_{c}^{2}+\left(2N_{c}^{2}-3\right)\left\langle [1,2,3,4]+[1,2,4,3]+[1,3,4,2]+[1,4,3,2]\right\rangle \nonumber \\
 & \left.-5N_{c}^{2}\left\langle [1][2,3,4]+[1][2,4,3]+[2][1,3,4]+[2][1,4,3]+[1,3][2,4]+[1,4][2,3]\right\rangle \frac{}{}\right]\nonumber \\
+ & \frac{\delta_{c_{2}'}^{c_{1}}\delta_{c_{1}'}^{c_{2}}\delta_{c_{4}'}^{c_{3}}\delta_{c_{3}'}^{c_{4}}}{\left(N_{c}^{2}\left(N_{c}^{2}-7\right)^{2}-36\right)}\left[\frac{}{}\left\langle [1,2][3,4]\right\rangle \left(N_{c}^{4}-8N_{c}^{2}+6\right)-\left(N_{c}^{2}-4\right)N_{c}^{2}\left\langle [1][2][3,4]+[1,2][3][4]\right\rangle \right.\nonumber \\
 & +\left(N_{c}^{2}+6\right)N_{c}^{2}\left\langle [1][2][3][4]\right\rangle -5N_{c}^{2}\left\langle [1][2,3][4]+[1][2,4][3]+[1,3][2][4]+[1,4][2][3]\right\rangle \nonumber \\
 & +\left(2N_{c}^{2}-3\right)\left(\left\langle [2,3,4][1]+[2,4,3][1]+[1,2,3][4]+[1,2,4][3]+[1,3,2][4]+[1,3,4][2]+[1,4,2][3]+[1,4,3][2]\right\rangle \right)\nonumber \\
 & +\left(4-N_{c}^{2}\right)\left\langle [1,2,3,4]+[1,2,4,3]+[1,3,4,2]+[1,4,3,2]\right\rangle +\left(N_{c}^{2}+6\right)\left\langle [1,3][2,4]+[1,4][2,3]\right\rangle \nonumber \\
 & \left.-5\left\langle [1,3,2,4]+[1,4,2,3]\right\rangle \frac{}{}\right]\nonumber \\
+ & \frac{\delta_{c_{2}'}^{c_{1}}\delta_{c_{3}'}^{c_{2}}\delta_{c_{1}'}^{c_{3}}\delta_{c_{4}'}^{c_{4}}}{\left(N_{c}^{2}\left(N_{c}^{2}-7\right)^{2}-36\right)}\left[\left(2N_{c}^{2}-3\right)N_{c}^{2}\left\langle [1][2][3][4]\right\rangle +\left(4-N_{c}^{2}\right)N_{c}^{2}\left\langle [1][2,3][4]+[1,2][3][4]+[1,3][2][4]\right\rangle \frac{}{}\right.\nonumber \\
 & +\left(N_{c}^{4}-8N_{c}^{2}+6\right)\left\langle [1,2,3][4]\right\rangle -5N_{c}^{2}\left\langle [1][2][3,4]+[1][2,4][3]+[1,4][2][3]\right\rangle \nonumber \\
 & +\left(2N_{c}^{2}-3\right)\left(\left\langle [2,3,4][1]+[1,2,4][3]+[1,3,2][4]+[1,4,3][2]+[1,2][3,4]+[1,3][2,4]+[1,4][2,3]\right\rangle \right)\nonumber \\
 & +\left(N_{c}^{2}+6\right)\left\langle [2,4,3][1]+[1,3,4][2]+[1,4,2][3]\right\rangle +\left(4-N_{c}^{2}\right)\left\langle [1,2,3,4]+[1,2,4,3]+[1,4,2,3]\right\rangle \nonumber \\
 & \left.-5\left(\left\langle [1,3,2,4]+[1,3,4,2]+[1,4,3,2]\right\rangle \right)\frac{}{}\right]\nonumber \\
+ & \frac{\delta_{c_{2}'}^{c_{1}}\delta_{c_{3}'}^{c_{2}}\delta_{c_{4}'}^{c_{3}}\delta_{c_{1}'}^{c_{4}}}{N_{c}\left(N_{c}^{2}\left(N_{c}^{2}-7\right)^{2}-36\right)}\left[\frac{}{}-5N_{c}^{4}\left\langle [1][2][3][4]\right\rangle +\left(N_{c}^{4}-8N_{c}^{2}+6\right)\left\langle [1,2,3,4]\right\rangle +N_{c}^{2}\left(N_{c}^{2}+6\right)\left\langle [2,4][1][3]+[1,3][2][4]\right\rangle \right.\nonumber \\
 & +\left(2N_{c}^{2}-3\right)N_{c}^{2}\left\langle [1][2][3,4]+[1][2,3][4]+[1,2][3][4]+[1,4][2][3]\right\rangle \nonumber \\
 & +N_{c}^{2}\left(4-N_{c}^{2}\right)\left\langle [2,3,4][1]+[1,2,4][3]+[1,3,4][2]+[1,2,3][4]+[1,2][3,4]+[1,4][2,3]\right\rangle \nonumber \\
 & -5N_{c}^{2}\left(\left\langle [1][2,4,3]\right\rangle +\left\langle [1,3][2,4]\right\rangle +\left\langle [1,3,2][4]\right\rangle +\left\langle [1,4,2][3]\right\rangle +\left\langle [1,4,3][2]\right\rangle \right)\nonumber \\
 & \left.+\left(2N_{c}^{2}-3\right)\left(\left\langle [1,2,4,3]\right\rangle +\left\langle [1,3,2,4]\right\rangle +\left\langle [1,3,4,2]\right\rangle +\left\langle [1,4,2,3]\right\rangle \right)+\left(N_{c}^{2}+6\right)\left\langle [1,4,3,2]\right\rangle \frac{}{}\right]\nonumber \\
 & -5N_{c}^{2}\left\langle [1][2,4,3]+[1,3][2,4]+[1,3,2][4]+[1,4,2][3]+[1,4,3][2]\right\rangle \nonumber \\
 & \left.+\left(2N_{c}^{2}-3\right)\left\langle [1,2,4,3]+[1,3,2,4]+[1,3,4,2]+[1,4,2,3]\right\rangle +\left(N_{c}^{2}+6\right)\left\langle [1,4,3,2]\right\rangle \frac{}{}\right]\nonumber \\
+ & \frac{\delta_{c_{2}'}^{c_{1}}\delta_{c_{4}'}^{c_{2}}\delta_{c_{1}'}^{c_{3}}\delta_{c_{3}'}^{c_{4}}}{N_{c}\left(N_{c}^{2}\left(N_{c}^{2}-7\right)^{2}-36\right)}\left[-5N_{c}^{4}\left\langle [1][2][3][4]\right\rangle +\left(N_{c}^{2}+6\right)N_{c}^{2}\left\langle [1][2,3][4]+[1,4][2][3]\right\rangle +\left(N_{c}^{4}-8N_{c}^{2}+6\right)\left\langle [1,2,4,3]\right\rangle \frac{}{}\right.\nonumber \\
 & +\left(2N_{c}^{2}-3\right)N_{c}^{2}\left\langle [1][2][3,4]+[2,4][1][3]+[1,2][3][4]+[1,3][2][4]\right\rangle \nonumber \\
 & -5N_{c}^{2}\left\langle [1][2,3,4]+[1,3,2][4]+[1,3,4][2]+[1,4][2,3]+[1,4,2][3]\right\rangle \nonumber \\
 & +N_{c}^{2}\left(4-N_{c}^{2}\right)\left\langle [2,4,3][1]+[1,2][3,4]+[1,3][2,4]+[1,2,3][4]+[1,2,4][3]+[1,4,3][2]\right\rangle \nonumber \\
 & \left.+\left(2N_{c}^{2}-3\right)\left\langle [1,2,3,4]+[1,3,2,4]+[1,4,2,3]+[1,4,3,2]\right\rangle +\left(N_{c}^{2}+6\right)\left\langle [1,3,4,2]\right\rangle \frac{}{}\right]\nonumber \\
+ & \frac{\delta_{c_{2}'}^{c_{1}}\delta_{c_{4}'}^{c_{2}}\delta_{c_{3}'}^{c_{3}}\delta_{c_{1}'}^{c_{4}}}{\left(N_{c}^{2}\left(N_{c}^{2}-7\right)^{2}-36\right)}\left[\left\langle [1][2][3][4]\right\rangle \left(2N_{c}^{2}-3\right)N_{c}^{2}+\left(4-N_{c}^{2}\right)N_{c}^{2}\left\langle [1][2,4][3]+[1,2][3][4]+[1,4][2][3]\right\rangle \frac{}{}\right.\nonumber \\
 & +\left(N_{c}^{4}-8N_{c}^{2}+6\right)\left\langle [1,2,4][3]\right\rangle -5N_{c}^{2}\left\langle [1][2][3,4]+[1][2,3][4]+[1,3][2][4]\right\rangle \nonumber \\
 & +\left(N_{c}^{2}+6\right)\left\langle [2,3,4][1]+[1,3,2][4]+[1,4,3][2]\right\rangle +\left(4-N_{c}^{2}\right)\left\langle [1,2,3,4]+[1,2,4,3]+[1,3,2,4]\right\rangle \nonumber \\
 & +\left(2N_{c}^{2}-3\right)\left\langle [2,4,3][1]+[1,2][3,4]+[1,2,3][4]+[1,3][2,4]+[1,3,4][2]+[1,4][2,3]+[1,4,2][3]\right\rangle \nonumber \\
 & \left.-5\left\langle [1,3,4,2]+[1,4,2,3]+[1,4,3,2]\right\rangle \frac{}{}\right]\nonumber \\
+ & \frac{\delta_{c_{3}'}^{c_{1}}\delta_{c_{1}'}^{c_{2}}\delta_{c_{2}'}^{c_{3}}\delta_{c_{4}'}^{c_{4}}}{\left(N_{c}^{2}\left(N_{c}^{2}-7\right)^{2}-36\right)}\left[\left(2N_{c}^{2}-3\right)N_{c}^{2}\left\langle [1][2][3][4]\right\rangle +\left(4-N_{c}^{2}\right)N_{c}^{2}\left\langle [1][2,3][4]+[1,2][3][4]+[1,3][2][4]\right\rangle \frac{}{}\right.\nonumber \\
 & +\left(N_{c}^{4}-8N_{c}^{2}+6\right)\left\langle [1,3,2][4]\right\rangle -5N_{c}^{2}\left\langle [1][2][3,4]+[1][2,4][3]+[1,4][2][3]\right\rangle \nonumber \\
 & +\left(2N_{c}^{2}-3\right)\left\langle [2,4,3][1]+[1,2][3,4]+[1,2,3][4]+[1,4,2][3]+[1,3][2,4]+[1,3,4][2]+[1,4][2,3]\right\rangle \nonumber \\
 & +\left\langle [2,3,4][1]+[1,2,4][3]+[1,4,3][2]\right\rangle \left(N_{c}^{2}+6\right)+\left(4-N_{c}^{2}\right)\left\langle [1,3,2,4]+[1,3,4,2]+[1,4,3,2]\right\rangle \nonumber \\
 & \left.-5\left\langle [1,2,3,4]+[1,2,4,3]+[1,4,2,3]\right\rangle \frac{}{}\right]\nonumber \\
+ & \frac{\delta_{c_{3}'}^{c_{1}}\delta_{c_{1}'}^{c_{2}}\delta_{c_{4}'}^{c_{3}}\delta_{c_{2}'}^{c_{4}}}{N_{c}\left(N_{c}^{2}\left(N_{c}^{2}-7\right)^{2}-36\right)}\left[-5N_{c}^{4}\left\langle [1][2][3][4]\right\rangle +\left(N_{c}^{2}+6\right)N_{c}^{2}\left\langle [1][2,3][4]+[1,4][2][3]\right\rangle \frac{}{}\right.\nonumber \\
 & +\left\langle [1,3,4,2]\right\rangle \left(N_{c}^{4}-8N_{c}^{2}+6\right)+\left(2N_{c}^{2}-3\right)N_{c}^{2}\left\langle [1][2][3,4]+[2,4][1][3]+[1,2][3][4]+[1,3][2][4]\right\rangle \nonumber \\
 & +N_{c}^{2}\left(4-N_{c}^{2}\right)\left\langle [2,3,4][1]+[1,2][3,4]+[1,3][2,4]+[1,3,2][4]+[1,3,4][2]+[1,4,2][3]\right\rangle \nonumber \\
 & -5N_{c}^{2}\left\langle [1][2,4,3]+[1,2,3][4]+[1,2,4][3]+[1,4][2,3]+[1,4,3][2]\right\rangle \nonumber \\
 & \left.+\left(2N_{c}^{2}-3\right)\left\langle [1,2,3,4]+[1,3,2,4]+[1,4,2,3]+[1,4,3,2]\right\rangle +\left(N_{c}^{2}+6\right)\left\langle [1,2,4,3]\right\rangle \frac{}{}\right]\nonumber \\
+ & \frac{\delta_{c_{3}'}^{c_{1}}\delta_{c_{2}'}^{c_{2}}\delta_{c_{1}'}^{c_{3}}\delta_{c_{4}'}^{c_{4}}}{N_{c}\left(N_{c}^{2}\left(N_{c}^{2}-7\right)^{2}-36\right)}\left[\frac{}{}\left(4-N_{c}^{2}\right)N_{c}^{4}\left\langle [1][2][3][4]\right\rangle +\left(N_{c}^{4}-8N_{c}^{2}+6\right)N_{c}^{2}\left\langle [1,3][2][4]\right\rangle \right.\nonumber \\
 & +\left\langle [2,4][1][3]\right\rangle \left(N_{c}^{2}+6\right)N_{c}^{2}+\left(2N_{c}^{2}-3\right)N_{c}^{2}\left\langle [1][2][3,4]+[1][2,3][4]+[1,2][3][4]+[1,4][2][3]\right\rangle \nonumber \\
 & +N_{c}^{2}\left(4-N_{c}^{2}\right)\left\langle [1,2,3][4]+[1,3][2,4]+[1,3,2][4]+[1,3,4][2]+[1,4,3][2]\right\rangle \nonumber \\
 & -5N_{c}^{2}\left\langle [1][2,3,4]+[1][2,4,3]+[1,2,4][3]+[1,4,2][3]+[1,2][3,4]+[1,4][2,3]\right\rangle \nonumber \\
 & \left.+\left(N_{c}^{2}+6\right)\left\langle [1,2,3,4]+[1,4,3,2]\right\rangle +\left(2N_{c}^{2}-3\right)\left\langle [1,2,4,3]+[1,3,2,4]+[1,3,4,2]+[1,4,2,3]\right\rangle \frac{}{}\right]\nonumber \\
+ & \frac{\delta_{c_{3}'}^{c_{1}}\delta_{c_{2}'}^{c_{2}}\delta_{c_{4}'}^{c_{3}}\delta_{c_{1}'}^{c_{4}}}{\left(N_{c}^{2}\left(N_{c}^{2}-7\right)^{2}-36\right)}\left[\left\langle [1][2][3][4]\right\rangle \left(2N_{c}^{2}-3\right)N_{c}^{2}+\left(N_{c}^{4}-8N_{c}^{2}+6\right)\left\langle [1,3,4][2]\right\rangle \frac{}{}\right.\nonumber \\
 & +\left(4-N_{c}^{2}\right)N_{c}^{2}\left\langle [1][2][3,4]+[1,3][2][4]+[1,4][2][3]\right\rangle -5N_{c}^{2}\left\langle [1][2,3][4]+[1][2,4][3]+[1,2][3][4]\right\rangle \nonumber \\
 & +\left(2N_{c}^{2}-3\right)\left\langle [2,3,4][1]+[1,2][3,4]+[1,2,4][3]+[1,3][2,4]+[1,3,2][4]+[1,4][2,3]+[1,4,3][2]\right\rangle \nonumber \\
 & +\left(N_{c}^{2}+6\right)\left\langle [2,4,3][1]+[1,2,3][4]+[1,4,2][3]\right\rangle +\left(4-N_{c}^{2}\right)\left\langle [1,2,3,4]+[1,3,2,4]+[1,3,4,2]\right\rangle \nonumber \\
 & \left.-5\left\langle [1,2,4,3]+[1,4,2,3]+[1,4,3,2]\right\rangle \frac{}{}\right]\nonumber \\
+ & \frac{\delta_{c_{3}'}^{c_{1}}\delta_{c_{4}'}^{c_{2}}\delta_{c_{1}'}^{c_{3}}\delta_{c_{2}'}^{c_{4}}}{\left(N_{c}^{2}\left(N_{c}^{2}-7\right)^{2}-36\right)}\left[\left\langle [1][2][3][4]\right\rangle N_{c}^{2}\left(N_{c}^{2}+6\right)+\left(N_{c}^{4}-8N_{c}^{2}+6\right)\left\langle [1,3][2,4]\right\rangle +N_{c}^{2}\left(4-N_{c}^{2}\right)\left\langle [2,4][1][3]+[1,3][2][4]\right\rangle \frac{}{}\right.\nonumber \\
 & +\left(2N_{c}^{2}-3\right)\left\langle [2,3,4][1]+[2,4,3][1]+[1,2,3][4]+[1,2,4][3]+[1,3,2][4]+[1,3,4][2]+[1,4,2][3]+[1,4,3][2]\right\rangle \nonumber \\
 & +\left(N_{c}^{2}+6\right)\left\langle [1,2][3,4]+[1,4][2,3]\right\rangle +\left(4-N_{c}^{2}\right)\left\langle [1,2,4,3]+[1,3,2,4]+[1,3,4,2]+[1,4,2,3]\right\rangle \nonumber \\
 & \left.-5N_{c}^{2}\left\langle [1][2][3,4]+[1][2,3][4]+[1,2][3][4]+[1,4][2][3]\right\rangle -5\left\langle [1,2,3,4]+[1,4,3,2]\right\rangle \frac{}{}\right]\nonumber \\
+ & \frac{\delta_{c_{3}'}^{c_{1}}\delta_{c_{4}'}^{c_{2}}\delta_{c_{2}'}^{c_{3}}\delta_{c_{1}'}^{c_{4}}}{N_{c}\left(N_{c}^{2}\left(N_{c}^{2}-7\right)^{2}-36\right)}\left[-5N_{c}^{4}[1][2][3][4]+\left(N_{c}^{4}-8N_{c}^{2}+6\right)\left\langle [1,3,2,4]\right\rangle +\left(N_{c}^{2}+6\right)N_{c}^{2}\left\langle [1][2][3,4]+[1,2][3][4]\right\rangle \frac{}{}\right.\nonumber \\
 & +\left(2N_{c}^{2}-3\right)N_{c}^{2}\left\langle [1][2,3][4]+[2,4][1][3]+[1,3][2][4]+[1,4][2][3]\right\rangle \nonumber \\
 & -5N_{c}^{2}\left\langle [1][2,3,4]+[1,2][3,4]+[1,2,3][4]+[1,4,2][3]+[1,4,3][2]\right\rangle \nonumber \\
 & +N_{c}^{2}\left(4-N_{c}^{2}\right)\left\langle [2,4,3][1]+[1,2,4][3]+[1,3][2,4]+[1,3,2][4]+[1,3,4][2]+[1,4][2,3]\right\rangle \nonumber \\
 & \left.+\left(2N_{c}^{2}-3\right)\left\langle [1,2,3,4]+[1,2,4,3]+[1,3,4,2]+[1,4,3,2]\right\rangle +\left(N_{c}^{2}+6\right)\left\langle [1,4,2,3]\right\rangle \frac{}{}\right]\nonumber \\
+ & \frac{\delta_{c_{4}'}^{c_{1}}\delta_{c_{1}'}^{c_{2}}\delta_{c_{2}'}^{c_{3}}\delta_{c_{3}'}^{c_{4}}}{N_{c}\left(N_{c}^{2}\left(N_{c}^{2}-7\right)^{2}-36\right)}\left[-5N_{c}^{4}\left\langle [1][2][3][4]\right\rangle +\left(2N_{c}^{2}-3\right)N_{c}^{2}\left\langle [1][2][3,4]+[1][2,3][4]+[1,2][3][4]+[1,4][2][3]\right\rangle \frac{}{}\right.\nonumber \\
 & +\left(N_{c}^{4}-8N_{c}^{2}+6\right)\left\langle [1,4,3,2]\right\rangle -5N_{c}^{2}\left\langle [1][2,3,4]+[1,2,3][4]+[1,2,4][3]+[1,3][2,4]+[1,3,4][2]\right\rangle \nonumber \\
 & +N_{c}^{2}\left(4-N_{c}^{2}\right)\left\langle [2,4,3][1]+[1,2][3,4]+[1,3,2][4]+[1,4][2,3]+[1,4,2][3]+[1,4,3][2]\right\rangle \nonumber \\
 & +N_{c}^{2}\left(N_{c}^{2}+6\right)\left\langle [2,4][1][3]+[1,3][2][4]\right\rangle +\left(2N_{c}^{2}-3\right)\left\langle [1,2,4,3]+[1,3,2,4]+[1,3,4,2]+[1,4,2,3]\right\rangle \nonumber \\
 & \left.+\left(N_{c}^{2}+6\right)\left\langle [1,2,3,4]\right\rangle \frac{}{}\right]\nonumber \\
+ & \frac{\delta_{c_{4}'}^{c_{1}}\delta_{c_{1}'}^{c_{2}}\delta_{c_{3}'}^{c_{3}}\delta_{c_{2}'}^{c_{4}}}{\left(N_{c}^{2}\left(N_{c}^{2}-7\right)^{2}-36\right)}\left[\left(2N_{c}^{2}-3\right)N_{c}^{2}\left\langle [1][2][3][4]\right\rangle +\left(N_{c}^{4}-8N_{c}^{2}+6\right)\left\langle [1,4,2][3]\right\rangle \frac{}{}\right.\nonumber \\
 & -5N_{c}^{2}\left\langle [1][2][3,4]+[1][2,3][4]+[1,3][2][4]\right\rangle \nonumber \\
 & +\left(2N_{c}^{2}-3\right)\left\langle [2,3,4][1]+[1,2][3,4]+[1,2,4][3]+[1,3][2,4]+[1,3,2][4]+[1,4][2,3]+[1,4,3][2]\right\rangle \nonumber \\
 & +N_{c}^{2}\left(4-N_{c}^{2}\right)\left\langle [1][2,4][3]+[1,2][3][4]+[1,4][2][3]\right\rangle +\left(4-N_{c}^{2}\right)\left\langle [1,3,4,2]+[1,4,2,3]+[1,4,3,2]\right\rangle \nonumber \\
 & \left.+\left(N_{c}^{2}+6\right)\left\langle [2,4,3][1]+[1,2,3][4]+[1,3,4][2]\right\rangle -5\left\langle [1,2,3,4]+[1,2,4,3]+[1,3,2,4]\right\rangle \frac{}{}\right]\nonumber \\
+ & \frac{\delta_{c_{4}'}^{c_{1}}\delta_{c_{2}'}^{c_{2}}\delta_{c_{1}'}^{c_{3}}\delta_{c_{3}'}^{c_{4}}}{\left(N_{c}^{2}\left(N_{c}^{2}-7\right)^{2}-36\right)}\left[\left\langle [1][2][3][4]\right\rangle \left(2N_{c}^{2}-3\right)N_{c}^{2}+\left(4-N_{c}^{2}\right)N_{c}^{2}\left\langle [1][2][3,4]+[1,3][2][4]+[1,4][2][3]\right\rangle \frac{}{}\right.\nonumber \\
 & +\left\langle [1,4,3][2]\right\rangle \left(N_{c}^{4}-8N_{c}^{2}+6\right)+\left(N_{c}^{2}+6\right)\left\langle [2,3,4][1]+[1,2,4][3]+[1,3,2][4]\right\rangle \nonumber \\
 & -5N_{c}^{2}\left\langle [1][2,3][4]+[1][2,4][3]+[1,2][3][4]\right\rangle +\left(4-N_{c}^{2}\right)\left\langle [1,2,4,3]+[1,3,4,2]+[1,4,2,3]+[1,4,3,2]\right\rangle \nonumber \\
 & +\left(2N_{c}^{2}-3\right)\left\langle [2,4,3][1]+[1,2][3,4]+[1,2,3][4]+[1,3][2,4]+[1,3,4][2]+[1,4][2,3]+[1,4,2][3]\right\rangle \nonumber \\
 & \left.-5\left\langle [1,2,3,4]+[1,3,2,4]\right\rangle \frac{}{}\right]\nonumber \\
+ & \frac{\delta_{c_{4}'}^{c_{1}}\delta_{c_{2}'}^{c_{2}}\delta_{c_{3}'}^{c_{3}}\delta_{c_{1}'}^{c_{4}}}{N_{c}\left(N_{c}^{2}\left(N_{c}^{2}-7\right)^{2}-36\right)}\left[\left(4-N_{c}^{2}\right)N_{c}^{4}\left\langle [1][2][3][4]\right\rangle +\left(N_{c}^{4}-8N_{c}^{2}+6\right)N_{c}^{2}\left\langle [1,4][2][3]\right\rangle \frac{}{}\right.\nonumber \\
 & +\left(2N_{c}^{2}-3\right)N_{c}^{2}\left\langle [1][2][3,4]+[2,4][1][3]+[1,2][3][4]+[1,3][2][4]\right\rangle +\left\langle [1][2,3][4]\right\rangle \left(N_{c}^{2}+6\right)N_{c}^{2}\nonumber \\
 & +\left(4-N_{c}^{2}\right)N_{c}^{2}\left\langle [1,2,4][3]+[1,3,4][2]+[1,4][2,3]+[1,4,2][3]+[1,4,3][2]\right\rangle \nonumber \\
 & -5N_{c}^{2}\left\langle [1][2,3,4]+[1][2,4,3]+[1,2][3,4]+[1,2,3][4]+[1,3][2,4]+[1,3,2][4]\right\rangle \nonumber \\
 & \left.+\left(2N_{c}^{2}-3\right)\left\langle [1,2,3,4]+[1,3,2,4]+[1,4,2,3]+[1,4,3,2]\right\rangle +\left(N_{c}^{2}+6\right)\left\langle [1,2,4,3]+[1,3,4,2]\right\rangle \frac{}{}\right]\nonumber \\
+ & \frac{\delta_{c_{4}'}^{c_{1}}\delta_{c_{3}'}^{c_{2}}\delta_{c_{1}'}^{c_{3}}\delta_{c_{2}'}^{c_{4}}}{N_{c}\left(N_{c}^{2}\left(N_{c}^{2}-7\right)^{2}-36\right)}\left[-5N_{c}^{4}\left\langle [1][2][3][4]\right\rangle +\left(N_{c}^{4}-8N_{c}^{2}+6\right)\left\langle [1,4,2,3]\right\rangle \frac{}{}\right.\nonumber \\
 & +\left(N_{c}^{2}+6\right)N_{c}^{2}\left\langle [1][2][3,4]+[1,2][3][4]\right\rangle +\left(2N_{c}^{2}-3\right)N_{c}^{2}\left\langle [1][2,3][4]+[2,4][1][3]+[1,3][2][4]+[1,4][2][3]\right\rangle \nonumber \\
 & +N_{c}^{2}\left(4-N_{c}^{2}\right)\left(\left\langle [2,3,4][1]+[1,2,3][4]+[1,3][2,4]+[1,4][2,3]+[1,4,2][3]+[1,4,3][2]\right\rangle \right)\nonumber \\
 & -5N_{c}^{2}\left\langle [1][2,4,3]+[1,2][3,4]+[1,2,4][3]+[1,3,2][4]+[1,3,4][2]\right\rangle \nonumber \\
 & \left.+\left(2N_{c}^{2}-3\right)\left\langle [1,2,3,4]+[1,2,4,3]+[1,3,4,2]+[1,4,3,2]\right\rangle +\left(N_{c}^{2}+6\right)\left\langle [1,3,2,4]\right\rangle \frac{}{}\right]\nonumber \\
+ & \frac{\delta_{c_{4}'}^{c_{1}}\delta_{c_{3}'}^{c_{2}}\delta_{c_{2}'}^{c_{3}}\delta_{c_{1}'}^{c_{4}}}{\left(N_{c}^{2}\left(N_{c}^{2}-7\right)^{2}-36\right)}\left[\left\langle [1][2][3][4]\right\rangle \left(N_{c}^{2}+6\right)N_{c}^{2}+\left\langle [1,4][2,3]\right\rangle \left(N_{c}^{4}-8N_{c}^{2}+6\right)\frac{}{}\right.\nonumber \\
 & +\left(4-N_{c}^{2}\right)N_{c}^{2}\left\langle [1][2,3][4]+[1,4][2][3]\right\rangle -5N_{c}^{2}\left\langle [1][2][3,4]+[1][2,4][3]+[1,2][3][4]+[1,3][2][4]\right\rangle \nonumber \\
 & +\left(2N_{c}^{2}-3\right)\left\langle [2,3,4][1]+[2,4,3][1]+[1,2,3][4]+[1,2,4][3]+[1,3,2][4]+[1,3,4][2]+[1,4,2][3]+[1,4,3][2]\right\rangle \nonumber \\
 & +\left(N_{c}^{2}+6\right)\left\langle [1,2][3,4]+[1,3][2,4]\right\rangle +\left(4-N_{c}^{2}\right)\left\langle [1,2,3,4]+[1,3,2,4]+[1,4,2,3]+[1,4,3,2]\right\rangle \nonumber \\
 & \left.-5\left(\left\langle [1,2,4,3]+[1,3,4,2]\right\rangle \right)\frac{}{}\right]\nonumber 
\end{align}

We need to mention that each term in the sum~(\ref{eq:AmpFull})
after contraction with partonic amplitudes can contribute with
different factor $N_{c}^{n},\quad n\in\mathbb{N}$, for this reason
it might be incorrect to take the limit $N_{c}\to\infty$ directly
in~~(\ref{eq:AmpFull}). As can be seen from~Figure~\ref{fig:QuarkoniaPairCGC}
and~(\ref{eq:NDef}), for diagrams 5,6 we will need only the amplitude
convoluted with color group generators
\begin{equation}
\mathcal{A}=\left(t_{a_{1}}\right)_{c_{1}}^{c_{1}'}\left(t_{a_{1}}\right)_{c_{2}}^{c_{2}'}\left(t_{a_{1}}\right)_{c_{3}}^{c_{3}'}\left(t_{a_{1}}\right)_{c_{4}}^{c_{4}'}\mathcal{A}_{c_{1}'c_{2}'c_{3}'c_{4}'}^{c_{1}c_{2}c_{3}c_{4}}
\end{equation}
Using Fierz identity
\begin{equation}
\left(t_{a_{1}}\right)_{c_{1}}^{c_{1}'}\left(t_{a_{1}}\right)_{c_{2}}^{c_{2}'}=\delta_{c_{2}}^{c_{1}'}\delta_{c_{1}}^{c_{2}'}-\frac{1}{N_{c}}\delta_{c_{1}}^{c_{1}'}\delta_{c_{2}}^{c_{2}'}
\end{equation}
and performing convolutions with the help of FeynCalc, we may obtain
a surprisingly simple result,
\begin{align}
\mathcal{A} & =\left\langle \left([1,2]-[1][2]\right)\left([3,4]-[3][4]\right)\right\rangle =\\
 & =\left\langle \left(S_{4}\left(\boldsymbol{x}_{1},\boldsymbol{y}_{1},\boldsymbol{x}_{2},\boldsymbol{y}_{2}\right)-S_{2}\left(\boldsymbol{x}_{1},\boldsymbol{y}_{1}\right)S_{2}\left(\boldsymbol{x}_{2},\boldsymbol{y}_{2}\right)\right)\left(S_{4}\left(\boldsymbol{x}_{3},\boldsymbol{y}_{3},\boldsymbol{x}_{4},\boldsymbol{y}_{4}\right)-S_{2}\left(\boldsymbol{x}_{3},\boldsymbol{y}_{3}\right)S_{2}\left(\boldsymbol{x}_{4},\boldsymbol{y}_{4}\right)\right)\right\rangle ,\nonumber 
\end{align}
namely all the contributions of sextupoles and octupoles cancel in
the final result. This is an untrivial property of the photoproduction
by color singlet photon: such cancellation does not occur if we consider
\textit{hadro}production or production of color octet $\bar{Q}Q$
pairs. Note that the result of this appendix are exact, namely we
haven't used the large-$N_{c}$ limit so far.

 \end{document}